\documentstyle[onecolumn]{mn}
\def\beq{\begin{equation}}
\def\eeq{\end{equation}}
\def\bey{\begin{eqnarray}}
\def\eey{\end{eqnarray}}

\def\Max{\rm Max}
\def\Min{\rm Min}
\input epsf

\title[Deprojection of light distributions of nearby systems]
	{Deprojection of light distributions of nearby systems: perspective effect and non-uniqueness} 
\author[HongSheng Zhao]
	{HongSheng Zhao
	\thanks{Sterrewacht Leiden (hsz@strw.LeidenUniv.nl)}}
\pagerange{\pageref{firstpage}--\pageref{lastpage}}
\pubyear{1999}

\begin{document}
\maketitle
\label{firstpage}

\begin{abstract}

Deriving the 3-dimensional volume density distribution from a
2-dimensional light distribution of a system yields generally
non-unique results.  The case for nearby systems is studied, taking
into account the extra constraints from the perspective effect.  It is
shown analytically that a new form of non-uniqueness exists.  The
Phantom Spheroid (PS) for a nearby system preserves the intrinsic
mirror symmetry and projected asymmetry of the system while changing
the shape and the major-axis orientation of the system.  A family of
analytical models are given as functions of the distance ($D_0$) to
the object and the amount ($\gamma$) of the PS density superimposed.  
The range of the major axis angles is constrained
analytically by requiring a positive density everywhere.  These models
suggest that observations other than surface brightness maps are
required to lift the degeneracy in the major axis angle and axis ratio
of the central bar of the Milky Way.
\end{abstract}


\section{Introduction}

Deprojection of galaxies from the observed light distribution on the
sky plane to the intrinsic 3-dimensional volume luminosity
distribution is one of the basic problems of astronomy.  It is common
knowledge that the deprojected results are generally non-unique
because of the freedom of distributing stars along any line of sight.
The best example is that a round distribution in projection may
correspond to any intrinsically prolate object pointing towards us.
Likewise we cannot tell from observed elliptical isophotes whether the
object is an intrinsically oblate bulge or a triaxial bar if both are
edge-on and at infinity (Contopoulos 1956).

The perspective effect of nearby triaxial objects, however, does make
them appear different from an oblate object.  The best example for
this is the famous left-to-right asymmetry of the dereddened light
distribution of the Milky Way bulge when plotted in the Galactic
$(l,b)$ coordinates, which led Blitz \& Spergel (1991) to conclude
that the Galactic bulge is in fact an almost edge-on bar, pointing at
an angle from the Sun.  They divide the Galaxy into a left and a right
part with the $l=0^o$ plane, which passes the Sun-center line and the
rotation axis of the Galaxy.  When folded along the $l=0^o$ line the
surface brightness map $I(l,b)$ is decomposed into two independent
maps: an asymmetry map $\left[I(l,b)-I(-l,b)\right]/2$ 
by subtracting the $l<0^o$ side from the $l>0^o$ side, and a symmetric map 
$\left[I(l,b)+I(-l,b)\right]/2$ by adding up the two sides.  
The signal in the asymmetry map,
they explain, is because the right hand side ($l>0^o$) of the bar
is nearer to us and 
the perspective effect makes it appear slightly bigger than the left hand side
($l<0^o$).  A simple sketch of the
geometry is shown in the top diagram of Fig.~\ref{cigar}.

The perspective effect allows Binney \& Gerhard (1996) and Binney,
Gerhard \& Spergel (1997) to derive a non-parametric volume density
distribution of the inner Galaxy.  The key element in their method is
to impose mirror symmetry for the bulge part of the Galaxy to allow
for a triaxial bar, and central symmetry for the disk part so to allow
for the spiral arm.  Although shifting material along the same line of
sight does not alter the isophotes, it spoils the symmetry of the
object.  Their numerical experiments suggest that the COBE map is
consistent with a range of bar models with the bar major axis pointing
some $15^o-35^o$ from the Sun.

Unfortunately there is a lack of analytical studies of general non-uniqueness
for nearby objects.  This is compared to a series of papers exploiting
the analytical properties of konuses in the deprojection of external
axisymmetric systems (Rybicki 1986, Palmer 1994, Kochanek \& Rybicki
1996, van den Bosch 1997, Romanowsky \& Kochanek 1997).  A small
amount of konuses, as christened by Gerhard \& Binney (1996) for a
well-studied class of artificial density models with zero surface
brightness, can be added to a galaxy at infinity without perturbing
its isophotes or creating negative-density zones.

This paper is a first attempt to give an analytical description of the
degeneracy in deprojecting the light distribution of a general nearby
system.  To ensure our arguments are not diluted by the necessary
mathematics, we will first introduce the so-called phantom spheroidal
(PS) density in \S2 and describe its effect on non-uniqueness in words
and illustrations.  The more mathematical aspects of the problem are
given in \S3-\S5, where we describe the properties of PS
densities and how to generate them.  We will then briefly discuss the
relations to the Milky Way bar in \S6, and relations with well-known
non-uniqueness of extragalactic objects in \S7.  We conclude in \S8.
Some additional results on the major axis angle and 
a generalization of the phantom density are given in the Appendix.

\section{Phantom Spheroid: a compromise between mirror symmetry and perspective effect}

\begin{figure}
\vskip -3cm
\epsfysize=12cm
\leftline{\epsfbox{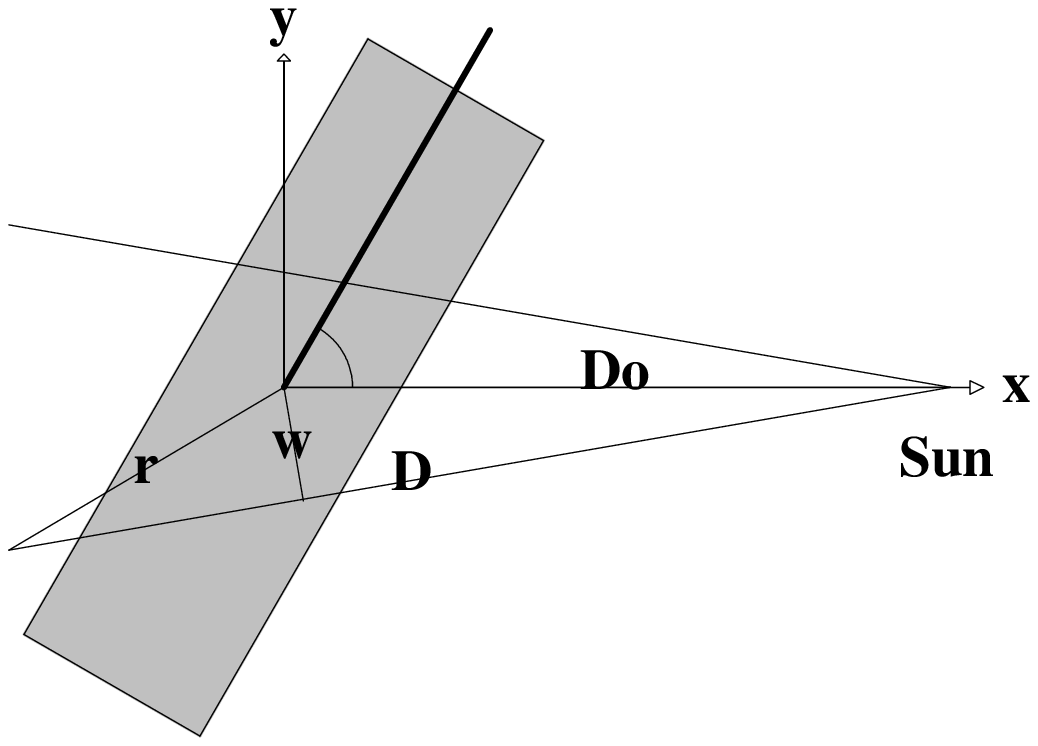}}
\vskip -7cm
\epsfysize=12cm
\leftline{\epsfbox{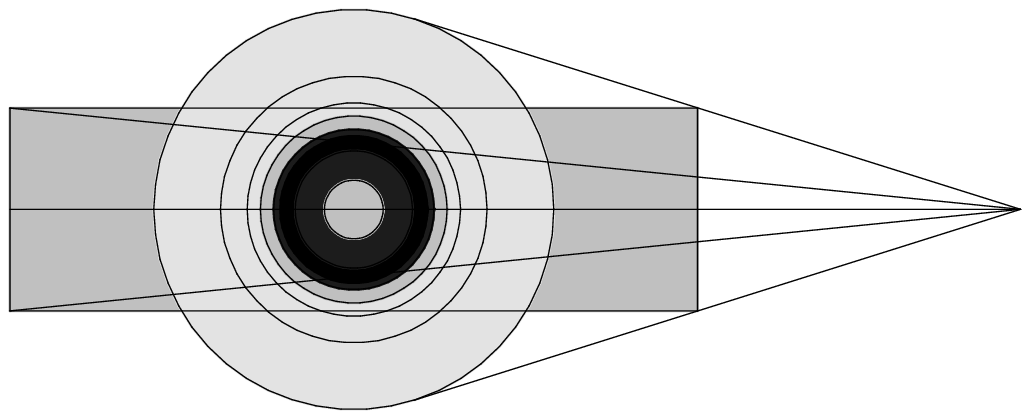}}
\vskip -7cm
\epsfysize=12cm
\leftline{\epsfbox{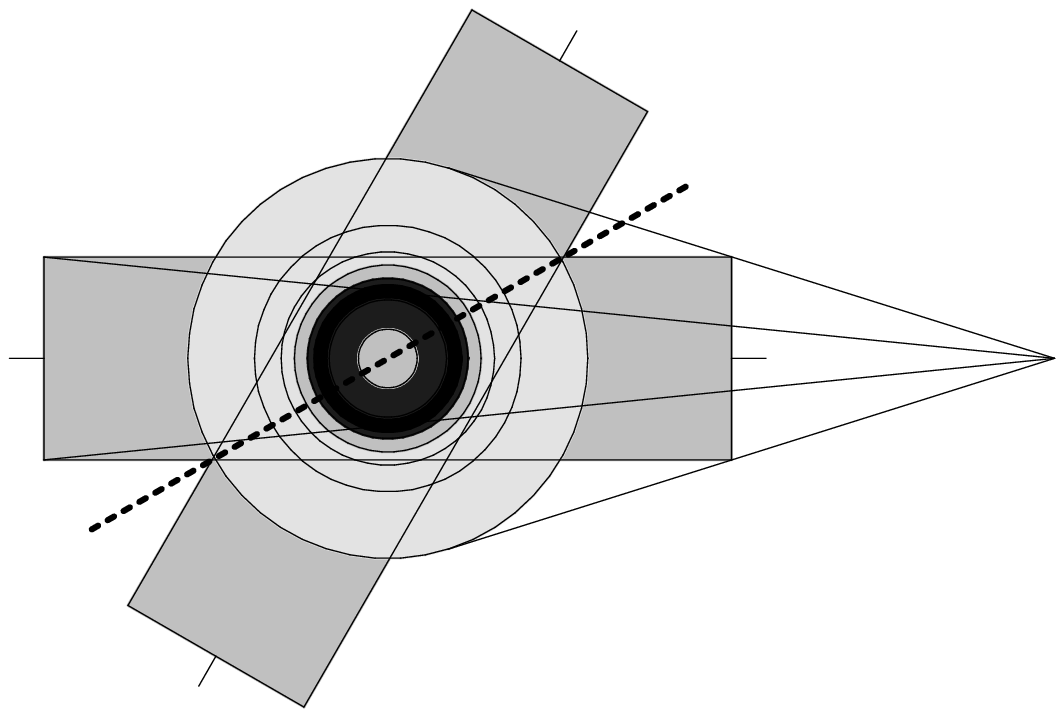}}
\vskip -3cm
\caption{
Top panel: the $xy$ cross section of a uniform cigar-shaped bar 
at distance $D_0$ with its major axis tilted at an angle $\alpha$ 
counterclockwise from the Sun-object center line.  The coordinates
of a point in the $xy$ plane can also be prescribed 
by its line of sight distance $D$, the impact parameter $w$, 
and the distance $r$ to the object center.
Middle panel: a phantom spheroid (PS).
Bottom panel: a superposition of the above two bodies.
Several line-of-sight pathes from the Sun
are drawn in thin solid lines.  The PS is made by
putting a clone of the cigar-shaped bar to an end-on geometry,
and subtracting from it a stratified ball-shaped bulge of
the same angular size and projected intensity.  
Adding such a PS to the cigar-shaped bar in the top diagram
has no effect on the latter's projected intensity, but rotates
the mirror plane of the spatial distribution 
to the dotted line in the middle at an angle ${\alpha \over 2}$
from the Sun-center line.}\label{cigar}
\end{figure}

A phantom density is a model with both positive and negative
density regions and a zero net surface brightness to an observer at
some distance away.  An example of a phantom density is illustrated
in the middle diagram of Fig.~\ref{cigar}, where we tailor the radial
profile of a stratified ball-shaped bulge in such a way so to have the
same angular size and projected intensity as a uniform cigar-shaped
bar placed end-on.  Subtracting the ball-shaped bulge from the
cigar-shaped bar yields a Phantom Spheroid (PS).  
The thin dark ring in the bulge is a density peak, corresponding to the line of
sight to the far edge of the cigar-shaped bar, a direction where the
depth and the projected intensity of the bar are at maximum.  The
fall-off of density towards the edge of the bulge corresponds to the
ever-decreasing depth of the cigar-shaped bar with increasing impact
parameter of the line of sight.

The problem with this simple cigar model is that if we superimpose
this unphysical component on any physical density distribution, say,
an ellipsoidal bar with a Gaussian radial profile, we get a new
density with generally twisted density patterns.  So while the added
cigar component is invisible from the observer's perspective, it
generally spoils the mirror symmetry of the system.

The exception is, as shown in the lower diagram of Fig.~\ref{cigar},
when the superimposed end-on cigar is a clone of the original cigar,
in which case the final density of the twin should have mirror symmetry with
respect to a new plane (the dotted line) which is just in between the
Sun-center line and the major axis of the original cigar.  In this
case the mirror symmetry is preserved, only the symmetry plane is
rotated.  Subtracting the round bulge has no further effect on the
symmetry plane, but will take away any trace of transformation in the
map of the integrated light.  Thus a PS which rotates the symmetry
axes and is invisible in projection can be obtained by placing the
original prolate cigar end-on and then subtracting off a round bulge.

We can apply the same trick to oblate systems, such as disks, because
a face-on disk mimics a spherical distribution the same way a prolate
cigar bar does.  So suppose the original model is a disk tilted at an
angle $\alpha$ from the line of sight.  Adding a face-on clone of the
disk will put the mirror plane of the twin disks at an angle ${\alpha
\over 2}$.  By subtracting off a spherical model we can take away any
effect of the face-on clone in the projected brightness map.  In fact
this kind of non-uniqueness applies any spheroidal (i.e. prolate or
oblate) distribution with any radial profile since our arguments about
non-uniqueness is independent of the aspect ratio and density profile
of the bar.  For example, the original model may consists of a small
spheroidal perturbation on top of a large positive spherical
component, both with general smooth radial profiles.

Despite these generalizations, the models are restricted in the sense
that the models come only as {\it twins} with identical projected
brightness distribution.  This is compared to non-uniqueness in the
extragalactic systems, such as konuses, where we find a {\it family}
of models with indistinguishable surface brightness maps with a
tunable amount of konuses.  The models here are also likely to have
large zones of negative density because we subtract off a significant
amount of matter with a spherical distribution.  This problem can be
softened if our original model has a large positive smooth spherical
component to start with.

\section{Method for generating general Phantom Spheroids}

Let's first set up a rectangular coordinate system $(x,y,z)$ centered
on the object with the $x$ axis pointing towards the observer.  In
this coordinate system the observer is at $(D_0,0,0)$, where $D_0$ is
the observer's distance to the object center.  From the observer's
perspective the surface brightness map of the system is most conveniently
specified by two angles $(l',b')$, 
which are the equivalent of the Galactic coordinate system
with the $b'=z=0$ plane being the $xy$ plane
and the $l'=y=0$ plane passing through $z$-axis of the system.
In these coordinates the system center 
and anti-center are in the directions $(l',b')=(0,0)$ and $(\pi,0)$.
A point $(x,y,z)$ in rectangular coordinates can 
thus be specified by the distance
$D$ to the observer along the line of sight $(l',b')$ with
\beq\label{lbxyz}
x=D_0-D\cos b' \cos l',~~~y=D\cos b' \sin l',~~~z=D\sin b'.
\eeq
The impact parameter $w$ 
with respect to the object center is given by
\beq\label{wlb}
w = D_0 \sqrt{\left(\sin b'\right)^2+ \left(\cos b' \sin l'\right)^2 }
\eeq
for the line of sight $(l',b')$.
The distance $r$ to the center of the object at $(0,0,0)$ is given by
\beq\label{ruxyz}
r=\sqrt{x^2+y^2+z^2}=\sqrt{\Delta^2+w^2},~~~\Delta=D-\sqrt{D_0^2-w^2},
\eeq
where $\Delta$ is the offset distance from the tangent point.  
The rectangular coordinates can also be expressed in terms of $w$ and $D$ with
\beq\label{wdxyz}
x=D_0-D\sqrt{1-{w^2 \over D_0^2}},~~~\sqrt{y^2+z^2}={D\over D_0}w.
\eeq
\begin{figure}
\vskip -3.5cm
\epsfysize=12cm
\centerline{\epsfbox{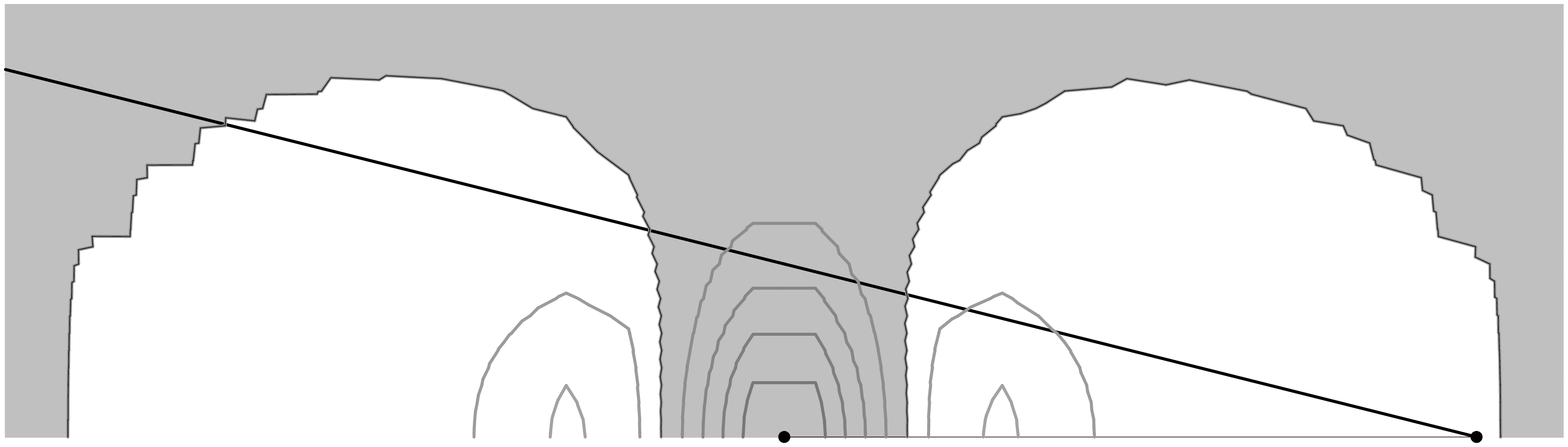}}
\vskip -6.5cm
\epsfysize=12cm
\centerline{\epsfbox{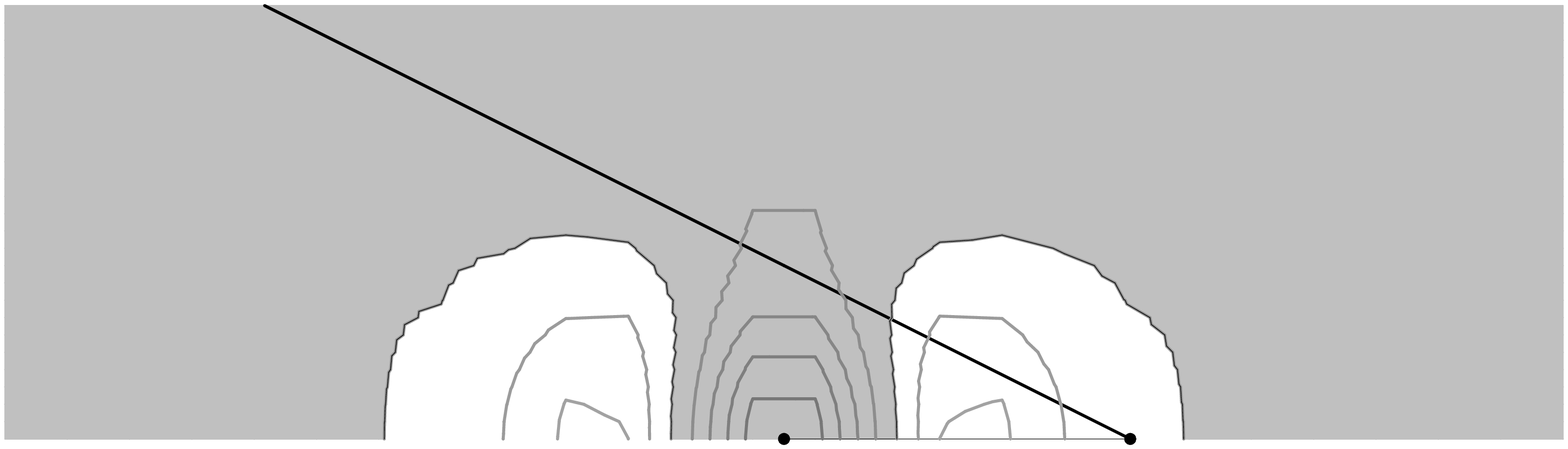}}
\vskip -6.5cm
\epsfysize=12cm
\centerline{\epsfbox{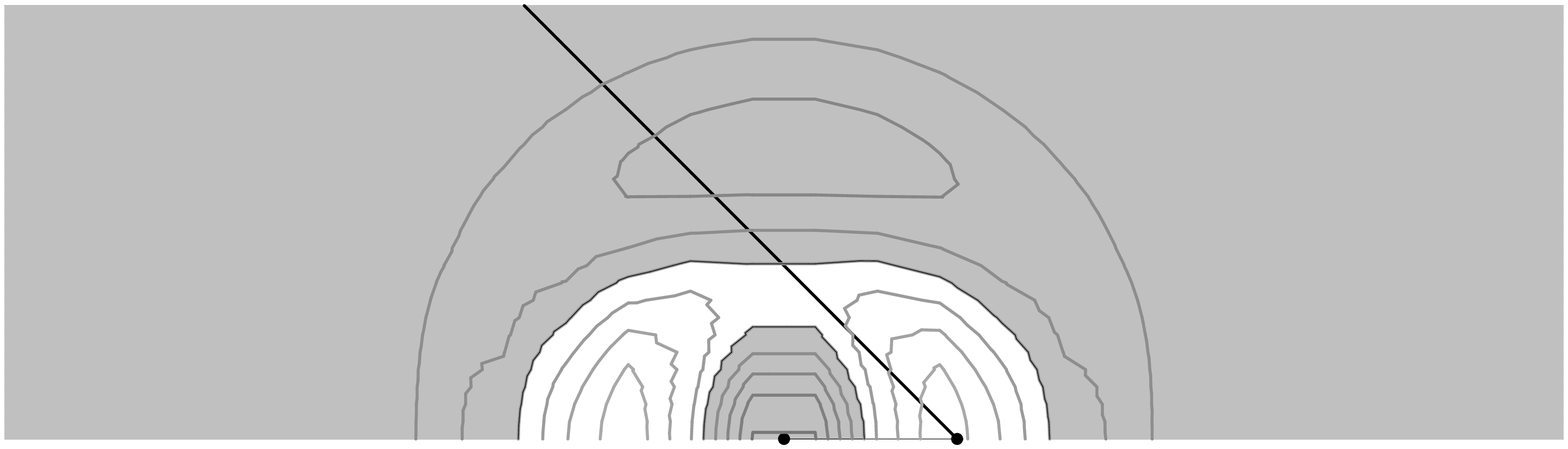}}
\vskip -6.5cm
\epsfysize=12cm
\centerline{\epsfbox{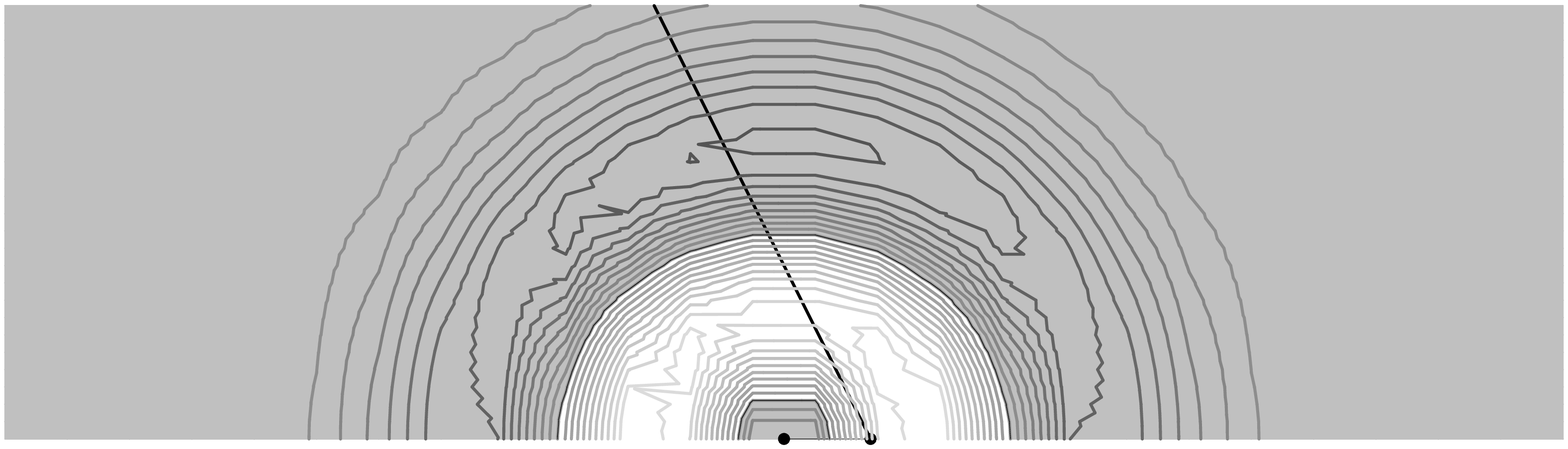}}
\vskip -3.5cm
\caption{
Cross sections of several phantom spheroidal density models
(cf. eq.~\protect{\ref{PS}}) in terms of $x$ 
(horizontal axis) vs. $\sqrt{y^2+z^2}$ (vertical axis).
The distributions are rotationally symmetric around the Sun-center axis
(i.e., the horizontal $x$-axis).
The shaded regions are negative density zones of the PS.
We show several locations of the observer (solid circle to the right)
at $D_0=4a, 2a, a, 0.5a$ (from top to bottom) from the object
center (solid circle to the lower left corner) and draw a line of sight
which intersects the $x=0$ plane at a radius $a$.  The amount of light 
adds up to zero along any line from the observer's perspective.
}\label{inv}
\end{figure}

Most generally our phantom spheroid $F(x,y,z)$ is
made by subtracting from an end-on
prolate distribution $P(|x|,\sqrt{y^2+z^2})$ 
a spherical bulge $S(r)$ with matching surface brightness, i.e.,
\beq\label{FF}
\int_0^\infty \!\! F(x,y,z) dD =0,~~~F(x,y,z) \equiv 
P(|x|,\sqrt{r^2-x^2}=\sqrt{y^2+z^2})-S(r),
\eeq
where $\int_0^\infty \!\! dD$ is an integration along any
line-of-sight direction, say, $(l',b')$.  Clearly $F(x,y,z)$ is an even
function of $x$, $y$ and $z$ with
\beq\label{Ffold}
F(x,y,z)=F(x,-y,z)=F(x,y,-z)=F(x,-y,-z)=F(-x,y,z),
\eeq
and in fact has rotational symmetry around the $x$-axis.
The total luminosity of a phantom $L_{PS}$ is given by
\beq\label{LPS}
L_{PS}\equiv \int\!\! d^3{\bf r} F(x,y,z) = L_P-L_S,
~~~L_P\equiv \int\!\! d^3{\bf r} P(|x|,\sqrt{y^2+z^2}),
~~~L_S\equiv \int\!\! d^3{\bf r} S(r).
\eeq
As we will show later a phantom spheroid can have a net luminosity because 
the total luminosity of the prolate component $L_P$
is only approximately that of the spherical component $L_S$
with the difference being infinitely small when the observer
is infinitely far away from the object.

Let $J(w)$ be the light intensity of 
the prolate distribution $P(|x|,\sqrt{y^2+z^2})$ 
integrated over a line of sight with impact parameter $w$,
including both the forward direction and 
the backward direction, then
\beq\label{sd}
J(w) \equiv \int_{-\infty}^{\infty} \! \! P(|x|,\sqrt{y^2+z^2}) dD,
\eeq
where $x,y,z$ can be expressed in terms of $w$ and $D$ 
using eq.~(\ref{ruxyz}) and eq.~(\ref{wdxyz}).  
Once $J(w)$ is computed from the integration
of $P(|x|,\sqrt{y^2+z^2})$, 
our task is to find a spherical bulge $S(r)$ such that
\beq\label{sd1}
J(w)=\int_{-\infty}^{\infty} \! \! S(r) dD.
\eeq
This can be done using the Abel transformation
\beq\label{invert}
S(r)=-{1 \over \pi} \int_{r}^{\infty} 
\! \! {dJ(w) \over dw} {dw \over \sqrt{w^2-r^2}}.
\eeq
The inversion uses effectively a variation of the well-known Eddington formula 
for deprojecting a spherical system (cf. Binney \& Tremaine 1987).  
Thus we find a general expression
for the phantom spheroid (PS); there is no restriction on
the radial profile or the axis ratio 
of the prolate distribution $P(|x|,\sqrt{y^2+z^2})$,
and in fact it is allowed to be oblate.
Any PS can also lead to a family of PS densities because
new PS can be generated by applying a linear operator to
an old PS.  For example, suppose $F(x,y,z)$ has a free parameter $\beta$,
then ${\partial \over \partial \beta} F(x,y,z)$ is also a PS (cf. Appendix B).

To reformulate the results in the previous section about the ``twin
bars'' in mathematical terms, we define 
a new set of rectangular coordinates $(x_\alpha,y_\alpha,z_\alpha)$
which relate to the rectangular coordinates $(x,y,z)$ by a rotation
around the $z$-axis (i.e., in the $xy$ plane) by an angle $\alpha$ with 
\beq
(x_\alpha,y_\alpha,z_\alpha)=
(x\cos \alpha + y\sin \alpha,-x\sin \alpha + y\cos \alpha, z).
\eeq
The rotation transformation has the property that
\bey\label{x12}
x_\alpha^2 &=&
x_{\alpha/2}^2\cos^2 {\alpha \over 2}+y_{\alpha/2}^2\sin^2 {\alpha \over 2}
+2x_{\alpha/2}y_{\alpha/2} \sin {\alpha \over 2} \cos {\alpha \over 2},\\\nonumber
x^2 &=&
x_{\alpha/2}^2\cos^2 {\alpha \over 2}+y_{\alpha/2}^2\sin^2 {\alpha \over 2}
-2x_{\alpha/2}y_{\alpha/2} \sin {\alpha \over 2} \cos {\alpha \over 2}.
\eey
In these notations we let $P(|x_\alpha|,\sqrt{r^2-x_\alpha^2})$
denote any general prolate bar, made by 
rotating its end-on twin bar $P(|x|,\sqrt{r^2-x^2})$ to an angle $\alpha$
from the line of sight in the $xy$ plane.  
Then we can always design a triaxial object
\beq\label{T}
T(|x_{\alpha/2}|,|y_{\alpha/2}|,|z_{\alpha/2}|)
\equiv P(|x_\alpha|,\sqrt{r^2-x_\alpha^2})+P(|x|,\sqrt{r^2-x^2})-S(r),
\eeq
where $S(r)$ is given by eqs.~(\ref{sd}) and~(\ref{invert}) so
that the triaxial object $T(|x_{\alpha/2}|,|y_{\alpha/2}|,|z_{\alpha/2}|)$
has exactly the same surface brightness as the prolate bar 
$P(|x_\alpha|,\sqrt{y_\alpha^2+z_\alpha^2})$, i.e.,
\beq\label{PT}
\int_{-\infty}^{\infty} \! \! P(|x_\alpha|,\sqrt{y_\alpha^2+z_\alpha^2}) dD =
\int_{-\infty}^{\infty} \! \! T(|x_{\alpha/2}|,|y_{\alpha/2}|,|z_{\alpha/2}|) dD.
\eeq
To verify the mirror symmetries of the
distribution $T(|x_{\alpha/2}|,|y_{\alpha/2}|,|z_{\alpha/2}|)$, 
simply substitute eq.~(\ref{x12}) to eq.~(\ref{T}) and apply the transformation
$x_{\alpha/2} \rightarrow -x_{\alpha/2}$ or
$y_{\alpha/2} \rightarrow -y_{\alpha/2}$.  In both cases we have
$|x_\alpha| \rightarrow |x|$.  The mirror symmetry with 
$z_{\alpha/2}=0$ plane is obvious because 
$T(|x_{\alpha/2}|,|y_{\alpha/2}|,|z_{\alpha/2}|)$ is an even function of 
$z_{\alpha/2}$, which comes in only through the
$r^2=x_{\alpha/2}^2+y_{\alpha/2}^2+z_{\alpha/2}^2$ factor.
Note that we do not lose generality by choosing the rotation 
to be around the $z$-axis since there is no prefered direction
in the $yz$ plane of the prolate bar $P(|x|,\sqrt{y^2+z^2})$.

To summarize the main result here, we have shown that {\it any}
prolate bar has a triaxial counterpart with identical surface
brightness but {\it half} the viewing angle.  An immediate question is
to what extent we can build triaxial models with the viewing angle
varying {\it continuously}.

\section{A general sequence of triaxial models with identical surface brightness}

In the following we will construct a subclass of these phantom spheroids
which preserve the mirror symmetry for a continuous sequence
of fairly general triaxial models.  We do not know 
whether such constructions exist for all triaxial models, but it does exist if 
the triaxial density distribution $\nu_0(x,y,z)$ can be
decomposed to a spherical part $G(r)$ and a non-spherical part 
$p(r)Q_0(x,y,z)$, i.e.,
\beq 
\nu_0(x,y,z)= G(r) + p(r)  Q_0(x,y,z),
\eeq
where the only restriction is that 
$Q_0(x,y,z)$ is a quadratic function of rectangular coordinates, i.e.,
\beq\label{Q0}
Q_0 \equiv c_{11} x^2 + c_{22} y^2 + c_{33} z^2 
+ 2 c_{12} xy + 2 c_{23} yz + 2 c_{31} zx.
\eeq
There is complete freedom with the functional forms of 
$G(r)$ and $r^2p(r)$, which are the the radial profiles
of the spherical component and the non-spherical component respectively.
These models are also triaxial by construction because
surfaces of constant $Q_0$ are ellipsoidal isosurfaces.
The symmetry planes of the ellipsoid can be oriented
to any direction by changing the elements of the symmetric matrix
$c_{ij}$ where $c_{ij}=c_{ji}$ and the indices $i$ and $j$ are from 1 to 3.
The total luminosity of the model, $L_0$, is given by
\beq\label{L0}
L_0\equiv \int\!\! d^3{\bf r} \nu_0(x,y,z) = 4\pi \int_0^\infty \! \! dr r^2 
G(r) + {4\pi \over 3} \sum_{i=1}^{3}c_{ii} \int_0^\infty \! \! dr r^4p(r).
\eeq
We define $I(l',b')$ as the surface brightness integrated
along both the $(l',b')$ direction and the opposite $(\pi+l',-b')$ direction.
The central surface brightness, $I_0$, integrated along the $x$-axis 
(i.e. with $x$ from $-\infty$ to $\infty$ and $y=z=0$) is given by
\beq\label{i01}
I_0 \equiv \int_{-\infty}^{\!\infty} \!\!\nu_0(x,0,0) dx
=\int_{-\infty}^{\!\infty} \!\!\left[G(r)+c_{11}r^2p(r)\right]dr,~~~r=x,
\eeq
where $r$ and $x$ are interchangeable dummy variables for the integrations.
\begin{figure}
\vskip -3cm
\epsfysize=12cm
\centerline{\epsfbox{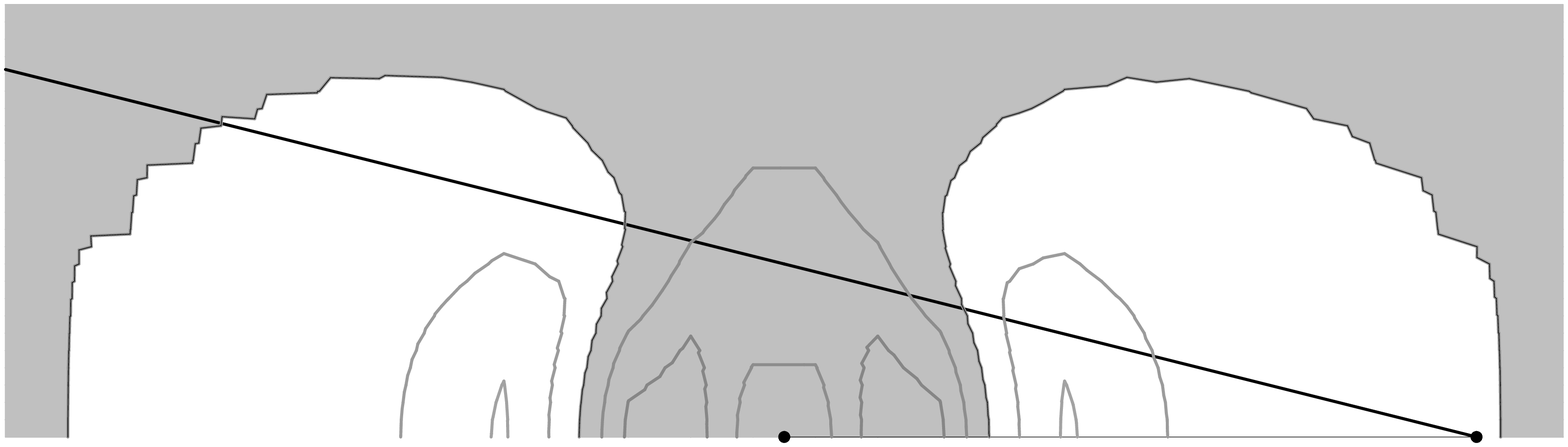}}
\vskip -6.5cm
\epsfysize=12cm
\centerline{\epsfbox{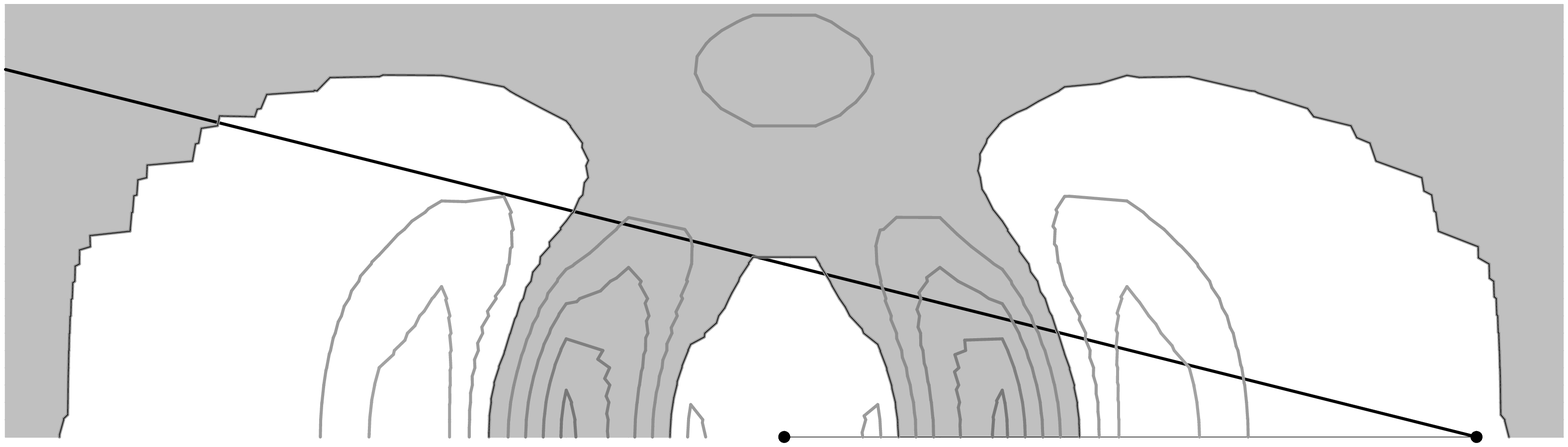}}
\vskip -6.5cm
\epsfysize=12cm
\centerline{\epsfbox{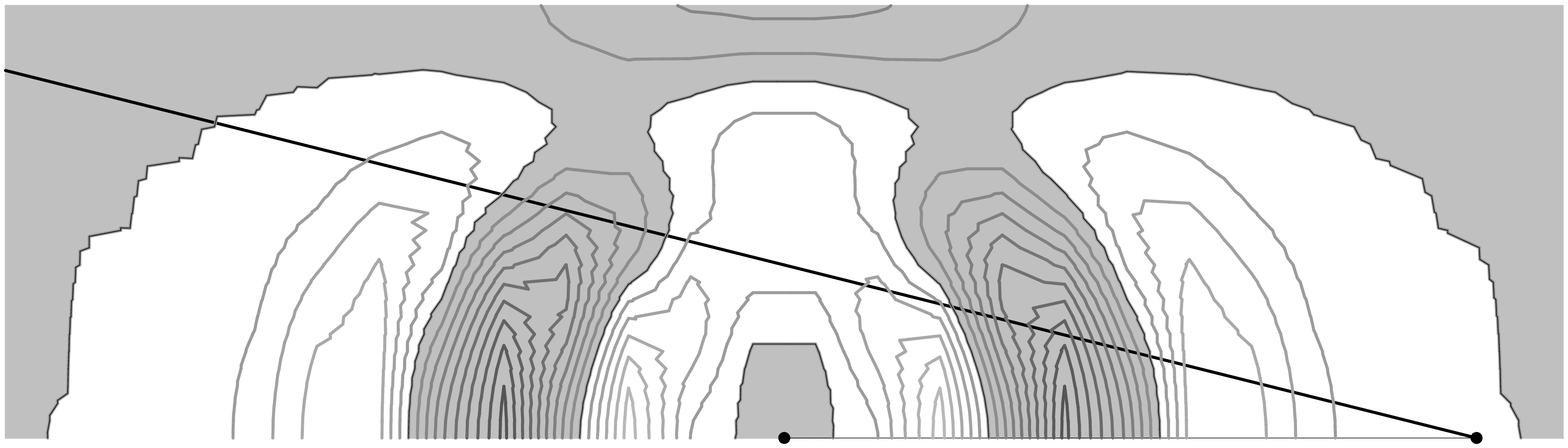}}
\vskip -3cm
\caption{
Same as the previous figure but 
for the new phantom spheroidal densities $F_1$, $F_2$ and $F_3$.
(cf. eq.~\protect{\ref{PSNEW}}) with the observer at $D_0=4a$.  
}\label{invnew}
\end{figure}

For the above triaxial model we can construct a phantom spheroid $F(x,y,z)$ with
\beq\label{f2}
F(x,y,z)= P(|x|,\sqrt{y^2+z^2}) -S(r),~~~ P(|x|,\sqrt{y^2+z^2}) = x^2p(r), 
\eeq
where the spherical component $S(r)$ is computed from (cf. eq.~\ref{invert})
\beq\label{invert2}
S(r)=-{1 \over \pi} \int_{r}^{\infty} 
\! \! {dJ(w) \over dw} {dw \over \sqrt{w^2-r^2}},
~~~J(w) =\int_{-\infty}^{\infty} \! \! x^2p(r) dD,
\eeq
so to cancel the $P(|x|,\sqrt{y^2+z^2})$ component 
in the projected light distribution.
Clearly $F(x,y,z)$ is a spheroidal distribution with
axial symmetry around the Sun-center axis since it is made
by subtracting from a prolate distribution $x^2p(r)$ 
a spherical distribution $S(r)$ of identical surface brightness.
Let $J_0$ be the line of sight integration of either the spherical $S(r)$ or
the prolate $x^2p(r)$ distribution along the $x$-axis, then
\beq\label{j01}
J_0 \equiv J(0)=\int_{-\infty}^{\!\infty} \!\!S(r)dr
=\int_{-\infty}^{\!\infty} \!\!r^2p(r)dr.
\eeq
The total luminosity of the phantom spheroid, $L_{PS}$, is given by
\beq\label{LPS2}
L_{PS}= L_P-L_S,
~~~L_P={4\pi \over 3}\int_0^\infty \! \! dr r^4p(r),
~~~L_S=4\pi \int_0^\infty \! \! dr r^2 S(r).
\eeq

Adding a fraction $\gamma$ of the above phantom spheroid 
to our original model $\nu_0(x,y,z)$ 
we get a new model
\beq\label{nu}
\nu_\gamma(x,y,z)=\nu_0(x,y,z) + F(x,y,z) \gamma,
\eeq 
which should have identical surface brightness map as the old one.
Rearranging the terms we find
\beq\label{nu1} 
\nu_\gamma(x,y,z) = G_\gamma(r) + p(r) Q_\gamma(x,y,z),
\eeq
where
\beq
G_\gamma \equiv G(r)-\gamma S(r),~~~r=\sqrt{x^2+y^2+z^2},
\eeq
and
\beq\label{Qu}
Q_\gamma(x,y,z) \equiv Q_0 (x,y,z) + \gamma x^2=
(\gamma+c_{11}) x^2 + c_{22} y^2 + c_{33} z^2 
+ 2 c_{12} xy + 2 c_{23} yz + 2 c_{31} zx.
\eeq
So the new model is very similar to the old model.
Both are superpositions of 
a spherical component and a triaxial perturbation;
the triaxiality is guaranteed by the fact that $Q_\gamma(x,y,z)$ prescribes
ellipsoidal isosurfaces.
In general the orientation of the object is a function of $\gamma$ given by
\beq\label{alphaxyz}
\cot 2\alpha_{xy} = {\gamma+c_{11}-c_{22} \over 2c_{12} },~~~
\cot 2\alpha_{xz} = {\gamma+c_{11}-c_{33} \over 2c_{13} },~~~
\cot 2\alpha_{yz} = {c_{22}-c_{33} \over 2c_{23} },
\eeq
where $\alpha_{xy}$ and 
$\alpha_{xz}$ are the tilt angles of the object from the 
the line of sight in the $xy$ plane (i.e., the $z=0$ cut)
and the $xz$ plane (i.e. the $y=0$ cut), 
and $\alpha_{yz}$ is the tilt angle counting from the $y$-axis in the
$yz$ plane (i.e., the $x=0$ cut).  We make two comments here.
First the orientation in the $yz$ plane is not affected by 
the phantom spheroid because the latter is generally rotationally symmetric
around in the $yz$ plane (cf. eq.~\ref{f2}).  Second the directions of
the principal axes of the triaxial object should generally be
eigenvectors of the $3\times 3$ matrix
$c_{ij}+\gamma \delta_{11}$ with $i,j=1,2,3$.  The corresponding eigenvalues
are generally the three roots of a cubic equation.  In practice, 
it is more convenient to speak of the orientation of the object
in terms of the angles $\alpha_{xy}$ and $\alpha_{xz}$, which 
prescribe the major axes in the $xy$ and $xz$ plane cuts.

There are, however, limits to $\gamma$, the amount of PS that we can add to
a model.  A physical density model must have a positive phase space
density and must be stable.  This, in principle, set limits on the
non-uniqueness if we can build dynamical (numerical) models for a
sequence of potentials with different $\gamma$.  In practice stringent
limits can already be obtained from the minimal requirement that the
volume density of a physical model is everywhere positive, i.e.,
\beq\label{nupos} 
\nu_\gamma(x,y,z) =\left[G(r)+p(r)Q_0(x,y,z)\right]+\gamma\left[x^2p(r)-S(r)\right]\ge 0.
\eeq
This generally involves (numerically) searching over a 3-dimensional space
$(x,y,z)$ for the minimum of the density $\nu_\gamma(x,y,z)$, and 
finding the range of $\gamma$ which brings the minimum above zero;
for edge-on models, this can be reduced to a search in 2-dimensional space
(see Appendix A).

Nevertheless there are many easy-to-use variations of the positivity 
equation.  For example, 
a positive volume density at the object center requires (cf. eq.~\ref{nupos})
\beq\label{cenden}
\nu_\gamma(0,0,0) = G_\gamma(0) = G(0) -\gamma S(0) \ge 0,
\eeq
while a less interesting condition can come from requiring 
a positive total luminosity of the object $L_\gamma$, given by
\beq\label{Lu}
L_\gamma\equiv \int\!\! d^3{\bf r} \nu_\gamma(x,y,z) = L_0+\gamma L_{PS} \ge 0.
\eeq
A stringent set of conditions can be derived by computing the following
moments of the density distribution.  This is done by
first multiplying both sides of eq.~(\ref{nupos}) by a factor $r^{k}$,
where $k$ can be 0, 1, 2 etc..
Then integrate over $r$ along a general direction ${\bf n}=(n_1, n_2, n_3)$
through the origin, i.e., from $r=-\infty$ to $r=\infty$ with
$(x,y,z)=(n_1 r,n_2 r,n_3 r)$.  We find
\beq
\int_{-\infty}^{\!\infty} \!\!\nu_\gamma(n_1 r,n_2 r,n_3 r) r^{k} dr=
\int_{-\infty}^{\!\infty} \!\!\left[G(r)+c_{nn}r^2p(r)\right] r^{k} dr
+\int_{-\infty}^{\!\infty} \!\!\left[\gamma n_1^2r^2p(r)-\gamma S(r)\right] r^{k} dr,
\eeq
where
\beq
n_1^2+n_2^2+n_3^2=1, ~~~c_{nn} \equiv \sum_{i,j=1}^{3} c_{ij}n_in_j.
\eeq

Requiring the $k=0$ moment of the density $\nu_\gamma(x,y,z)$ to be positive
we have
\beq\label{flim0}
\int_{-\infty}^{\!\infty} \!\!\nu_\gamma(n_1 r,n_2 r,n_3 r) dr= 
I_0 + 
\left[\left(\sum_{i,j=1}^{3} c_{ij}n_in_j-c_{11}\right)
-\left(n_2^2+n_3^2\right)\gamma\right] J_0 
\ge 0,
\eeq
where we have substituted in the definitions for $I_0$ and $J_0$ 
(cf. eqns.~\ref{i01} and~\ref{j01}).
Eq.~(\ref{flim0}) generally yields a necessary upper limit on $\gamma$, and
the limit is most stringent when the line $(x,y,z)=(n_1 r,n_2 r,n_3
r)$ is the minor axis of the model.  
For example, along the $y$-axis with $(n_1,n_2,n_3)=(0,1,0)$ we have 
\beq\label{alphaxylim}
2c_{12} \cot 2\alpha_{xy}=\gamma+c_{11}-c_{22} \le {I_0 \over J_0}
\eeq
where we express the constraint in terms of 
the tilt angle $\alpha_{xy}$ (cf. eq.~\ref{alphaxyz}) as well as $\gamma$.
Note we assume $J_0>0$, which is generally valid.  
Likewise along the $z$-axis with $(n_1,n_2,n_3)=(0,0,1)$ we obtain
\beq\label{alphaxzlim}
2c_{13} \cot 2\alpha_{xz}=\gamma+c_{11}-c_{33} \le {I_0 \over J_0}.
\eeq

Likewise a positive $k=2$ moment of the density requires
\beq\label{flim2}
4\pi \int_{-\infty}^{\!\infty} \!\!\nu_\gamma(n_1 r,n_2 r,n_3 r) r^2dr= 
L_0
+\left(3\sum_{i,j=1}^{3} c_{ij}n_in_j-\sum_{i=1}^{3} c_{ii} \right)L_P
+ \gamma (3 n_1^2 L_P - L_S) \ge 0,
\eeq
where we have substituted in the definitions for $L_0$, $L_P$ and $L_S$
(cf. eq.~\ref{L0} and~\ref{LPS2}).
Eq.~(\ref{flim2}) generally yields a necessary lower limit on $\gamma$.
For example, along the $x$-axis with ${\bf n}=(1,0,0)$ we have
\beq\label{xmass}
2c_{12} \cot 2\alpha_{xy} +
2c_{13} \cot 2\alpha_{xz} 
=2\gamma+2c_{11}-c_{22}-c_{33} \ge -{L_\gamma \over L_P},
\eeq
where we assume the prolate component has a positive luminosity,
i.e., $L_P>0$, which is generally the case.  The r.h.s of eq.~(\ref{xmass})
can be further simplied in the limit $D_0\rightarrow \infty$, i.e.,
the observer is far away from the object.  In this case $L_\gamma \rightarrow L_0$.

\begin{figure}
\epsfysize=6cm
\epsfxsize=15cm
\leftline{\epsfbox{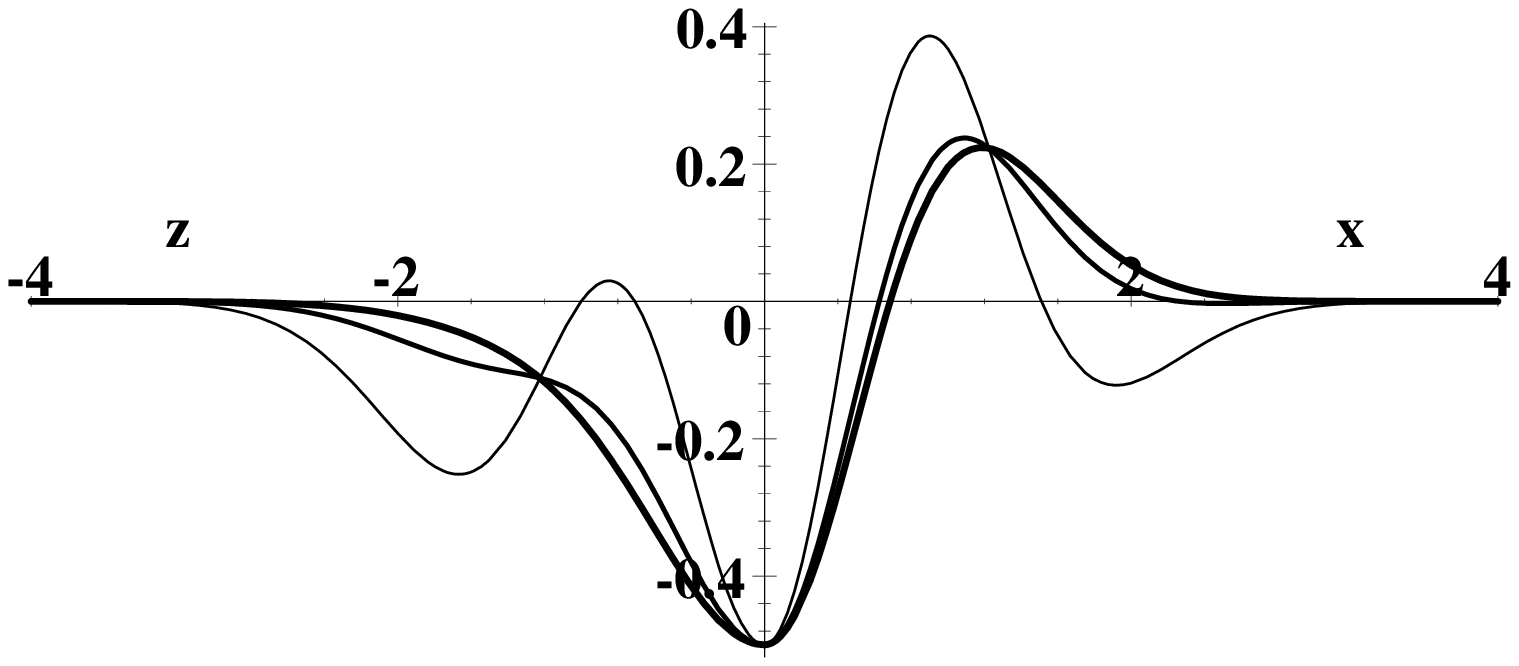}}
\vskip -1cm
\epsfysize=6cm
\epsfxsize=15cm
\leftline{\epsfbox{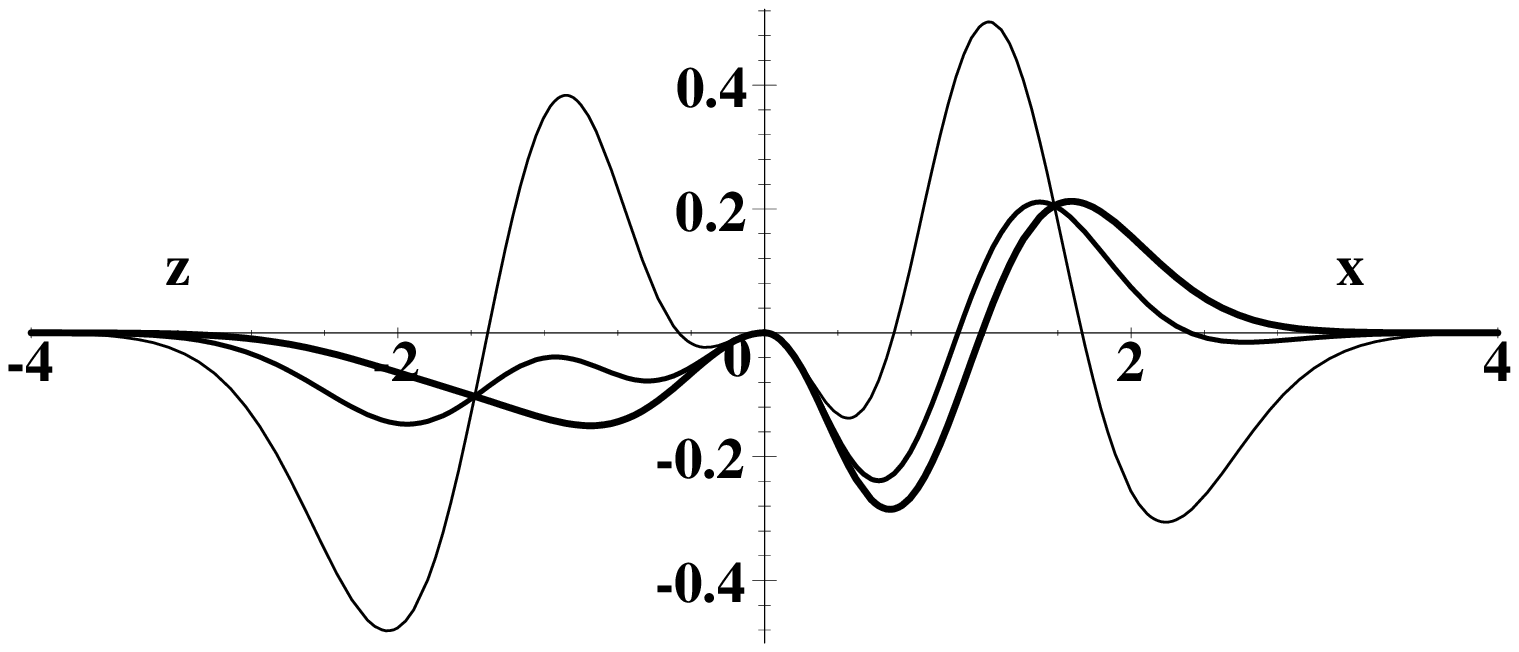}}
\vskip -1cm
\epsfysize=6cm
\epsfxsize=15cm
\leftline{\epsfbox{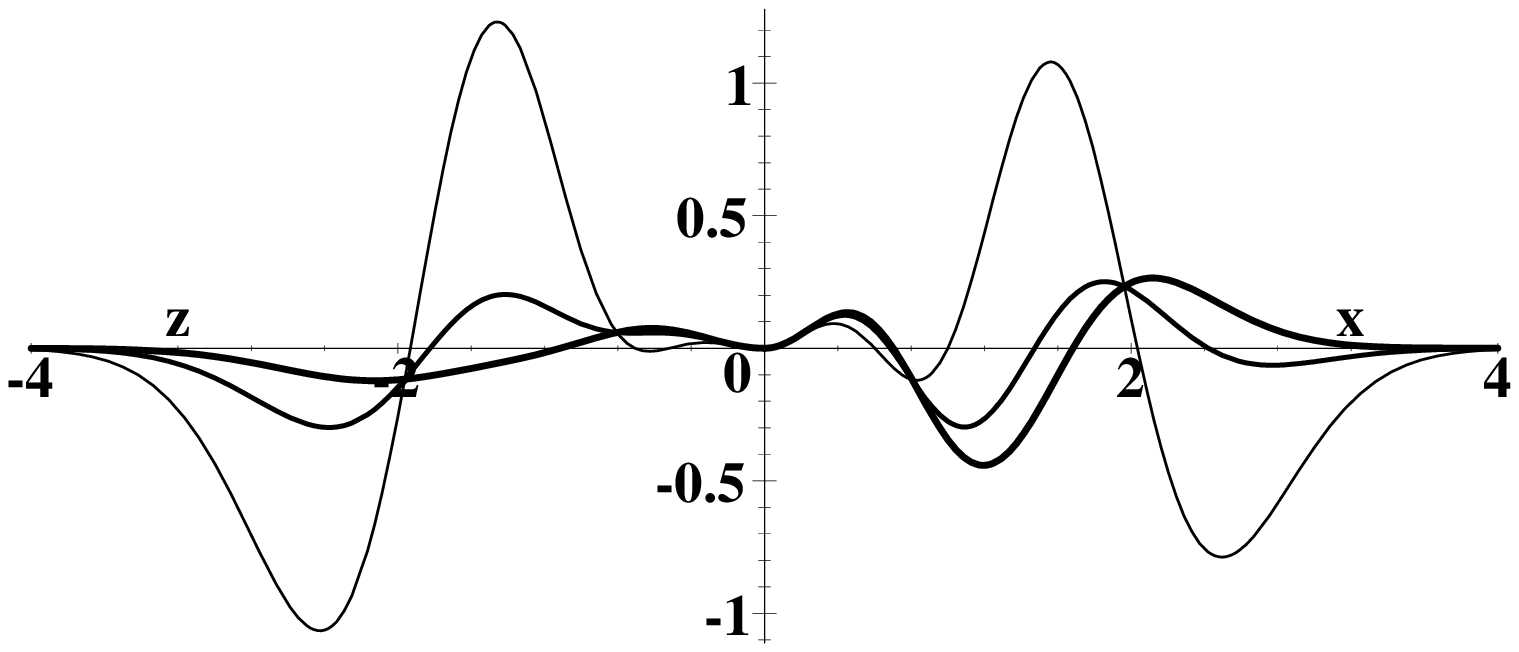}}
\caption{
The run of the phantom densities $F$ (top), $F_1$ (middle) and $F_2$ (bottom)
as a function of the radius, where the length 
is in units of the characteristic length $a$, 
and the density has units of $\nu_c$.  
The run along the $z$-axis is shown in the left half
(the negative half of the horizontal axis), 
and the run along the $x$-axis is shown in the
right half (positive half).  Various line styles
are for $D_0=4a$, $2a$ and $a$ with heavier lines for larger $D_0$.
}\label{invplts}
\end{figure}

\section{A sequence of nearly edge-on triaxial models with a Gaussian profile}

The above general results apply to models with any radial profile.
To illustrate the model properties effectively, we will use
the following smooth triaxial models with a Gaussian radial profile with
\beq\label{nu0a}
\nu_0(x,y,z)= G(r) + p(r)  Q_0(x,y,z),~~~G(r)=\nu_c e^{-{r^2 \over 2 a^2}},
~~~p(r)= {\nu_c \over a^2} e^{-{r^2 \over a^2}},
\eeq
where $a$ is the characteristic scale of the model,
$\nu_c$ is a characteristic density.  These two intrinsic parameters
are related to the observable $I_0$ (cf. eq.~\ref{i01}) by
\beq\label{i0}
I_0 =\int_{-\infty}^{\!\infty} \!\!\nu_0(x,0,0) dx
= \sqrt{\pi}\left(\sqrt{2}+{c_{11} \over 2}\right) \nu_c a,
\eeq
where $I_0$ is effectively the integrated light intensity from both the object
center $(l',b')=(0,0)$ and the anti-center $(l',b')=(\pi,0)$.
While we prefer to keep our derivations as general as possible,  
wherever quantitative numerical calculations are shown we 
set the scaling parameters $a$ and $\nu_c$ to unity,
and use the following set of model parameters
\beq\label{examp}
c_{11}=c_{22}=-{5 \over 6},~c_{33}=-{7 \over 6},~c_{12}={1 \over 3},
\eeq
and
\beq
c_{31}=c_{23}={1 \over 9},
\eeq
and put the object at the distance 
\beq
D_0=4a
\eeq
unless otherwise mentioned.  These parameters describe
nearly edge-on models with a characteristic angular 
size of ${a \over D_0}={1 \over 4} \sim 15^o$;
edge-on models have $c_{31}=c_{23}=0$.
We will vary $\gamma$ to show the properties of a sequence of models.
The results can be generalized qualitatively to all models described
by eq.~(\ref{nu1}).

To find the expression for the phantom spheroid, 
we substitute eq.~(\ref{nu0a}) in eq.~(\ref{sd}).  This yields
\beq\label{j0}
J(w)=\int_{-\infty}^{\infty} \! \! x^2p(r) dD 
= J_0  e^{-{w^2 \over a^2}}\left(
1 + {2w^2 \over a^2} - {w^2\over D_0^2} \right),
~~~J_0= {\sqrt{\pi} \over 2} \nu_c a,
\eeq
where $J(w)$ is the 
integrated intensity of the prolate distribution $x^2p(r)$
along the line of sight with the impact parameter $w$, 
which becomes $J_0$ when $w=0$.
Applying eq.~(\ref{invert}) we find that 
$S(r)$ reduces to a simple analytical expression,
\beq\label{sr}
S(r)=-{1 \over \pi} \int_{r}^{\infty} 
\! \! {dJ(w) \over dw} {dw \over \sqrt{w^2-r^2}}
= a^2p(r)\left( {1 \over 2} + 
{r^4 \over a^2 D_0^2} - {3 r^2\over 2 D_0^2} \right),
\eeq
and the PS is given by
\beq\label{PS}
F(x,y,z) = x^2p-S= \nu_c e^{-{r^2 \over a^2}} \left[ 
\left({x^2 \over a^2}-{1 \over 2}\right) + {a^2 \over D_0^2} 
\left({3r^2 \over 2a^2}-{r^4 \over a^4} \right) \right].
\eeq
A few examples of the PS are shown in Fig.~\ref{inv}.  
They vary as a function of characteristic angular size of the object
${a \over D_0}$.  When the observer is well inside the object, 
${a \over D_0} \gg 1$, the PS density becomes nearly spherical
because of the strong dependence on $r$ (cf. eq.~\ref{PS}).
The amplitude changes from $\sim -{a^2 \over 5 D_0^2}\nu_c$ 
to $\sim {a^2 \over 5 D_0^2}\nu_c$ when $r$ moves from $r=a$ to $r=1.5a$.

We can also generate a sequence of analytical phantom spheroids by repeatedly 
applying the derivative operator ${1 \over 2}{\partial \over \partial \ln a}$
to the original phantom spheroid $F(x,y,z)$ (cf. eq.~\ref{PS}), i.e,
\beq\label{PSNEW}
F_n(x,y,z)=x^2p_n(r)-S_n(r),
~~~p_n(r) \equiv {\partial^n p(r) \over 2^n \partial(\ln a)^n},
~~~S_n(r) \equiv {\partial^n S(r) \over 2^n \partial(\ln a)^n}
\eeq
for $n=1,2,3,...$, where $x^2p_n(r)$ and $S_n(r)$ form
a new pair of a prolate bar and a spherical bulge with matching
surface brightness.  These new analytical PS densities are given 
in the Appendix B.
The contours for these new PS densities
are shown in Fig.~\ref{invnew}, and radial profiles in Fig.~\ref{invplts}.
As $D_0$ decreases, the amplitude of the radial oscillation of the PS increases,
but at the origin we have, 
\beq
F(0,0,0)=-0.5\nu_c,~F_1(0,0,0)=F_2(0,0,0)=0,
\eeq
independent of $D_0$.  

\begin{figure}
\vskip -2cm
\epsfysize=10cm
\centerline{\epsfbox{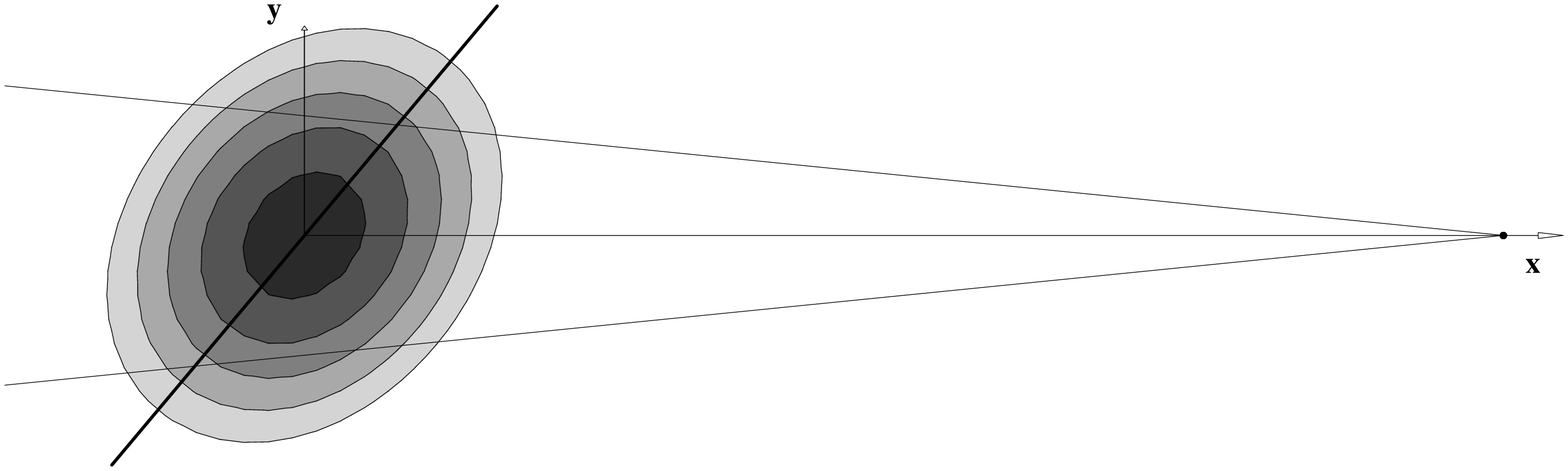}}
\vskip -5cm
\epsfysize=10cm
\centerline{\epsfbox{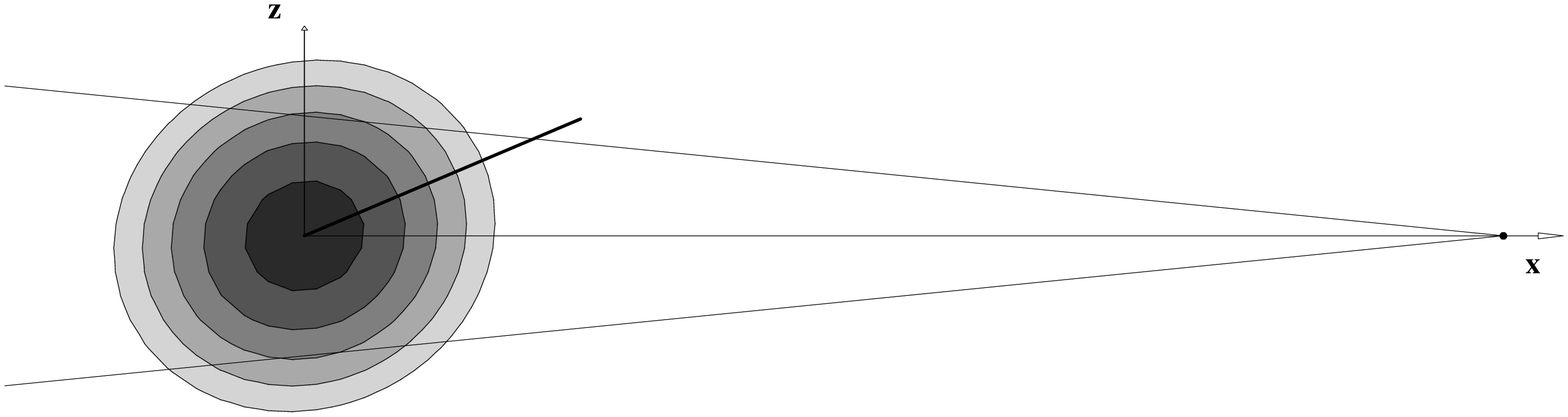}}
\vskip -2cm
\caption{
Cross sections of a triaxial nearly edge-on model 
(cf. eq.~\protect{\ref{examp}})
with the major axis at $\alpha_{xy}=50^o$ from the Sun-center line
in the $xy$ plane (top) and at $\alpha_{xz}=23^o$
in the $xz$ plane (bottom). 
}\label{bar1}
\vskip -2cm
\epsfysize=10cm
\centerline{\epsfbox{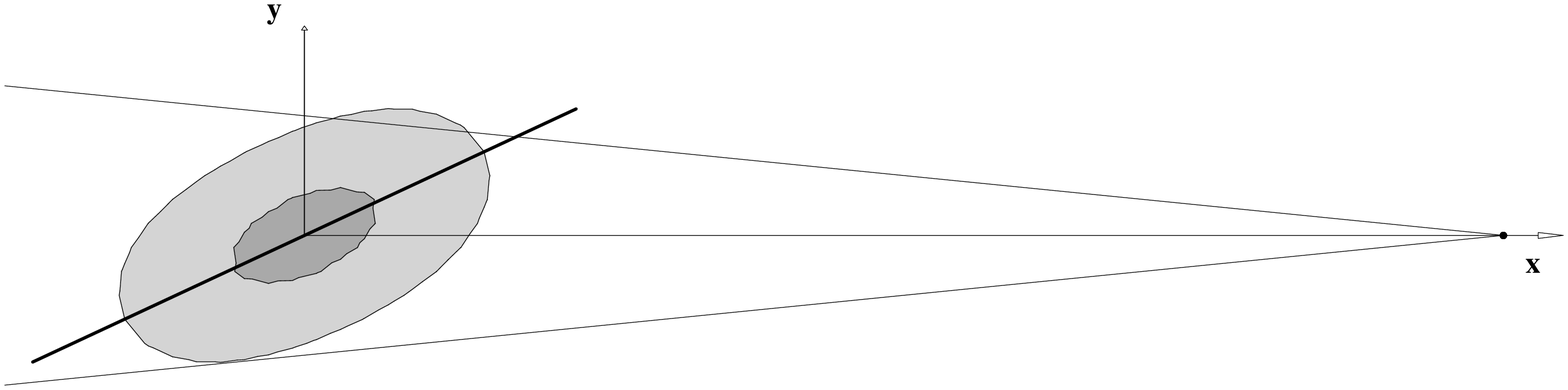}}
\vskip -5cm
\epsfysize=10cm
\centerline{\epsfbox{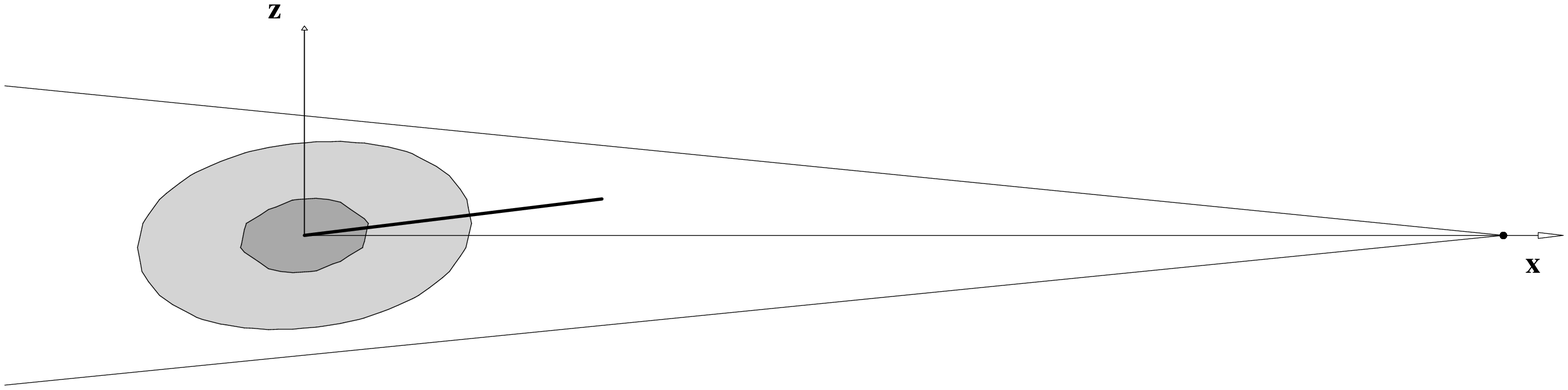}}
\vskip -2cm
\caption{
Similar to the previous diagram, but 
with the major axis at $\alpha_{xy}=25^o$ from the Sun-center line
in the $xy$ plane (top) and at $\alpha_{xz}=7^o$
in the $xz$ plane (bottom).
This model appears identical to the previous model in terms of
the surface brightness distribution.
}\label{bar2}
\end{figure}
Adding the phantom spheroidal density $F(x,y,z)$ to our $\gamma=0$ model
(cf. eq.~\ref{nu0a}), we obtain the general expression for the model density
\bey\label{nu2}
\nu_\gamma(x,y,z)&=& 
\nu_c e^{-{r^2 \over 2 a^2}} \left\{1
+ e^{-{r^2 \over 2a^2}} \left[{Q_\gamma \over a^2} -{\gamma \over 2} -
{\gamma r^4 \over a^2 D_0^2} + {3 \gamma r^2\over 2 D_0^2} \right]
\right\},\\\nonumber
Q_\gamma&=&(\gamma+c_{11}) x^2 + c_{22} y^2 + c_{33} z^2 
+ 2 c_{12} xy + 2 c_{23} yz + 2 c_{31} zx.
\eey
The object-observer distance $D_0$ can take on any positive value,
and the parameters $\gamma$ and $c_{ij}$ can describe the most 
general orientation of the three symmetry planes of the model 
(cf. eq.~\ref{Qu}).  These
models have also the nice property that the PS density falls
off steeper than the spherical term $G(r)$, meaning that the models are
always nearly spherical with a positive, Gaussian profile at large radii.

\begin{figure}
\epsfysize=10cm
\centerline{\epsfbox{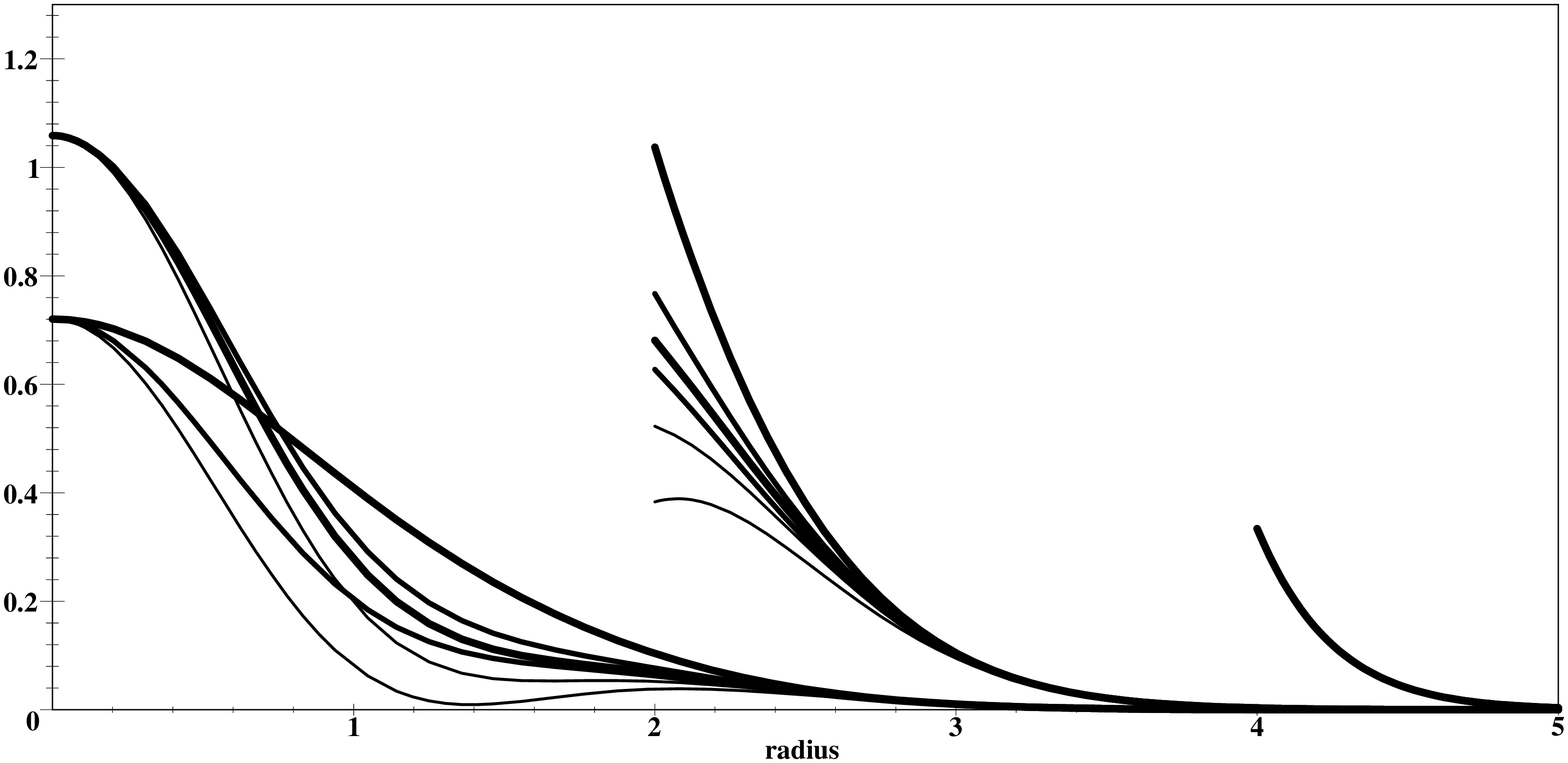}}
\caption{The run of the volume density along the $x$ (thickest line), $y$
and $z$ (thinnest line) axes for the $\alpha_{xy}=50^o$ model (upper curves)
and the $\alpha_{xy}=25^o$ model (lower curves). 
The curves  are blown up at large radius 
by a factor of 10 and 1000, and they converge to one curve 
because the density at large radius is dominated by the spherical Gaussian 
component (cf. eq.~\ref{nu2}).
}\label{barxyz}
\end{figure}

The free parameter $\gamma$ is constrained by 
the requirement of a positive volume density to the range
\beq
\Min(\gamma) < \gamma < \Max(\gamma),~~~\Min(\gamma) \sim -0.5,~~~\Max(\gamma) \sim 0.6,
\eeq
where the lower and upper limits are found by computing numerically the density
on a spatial grid in the $(x,y,z)$ and examining the positivity
(cf. eq.~\ref{nupos} with the parameters in eq.~\ref{examp}).  
They translates to 
\beq
\Min(\alpha_{xy}) < \alpha_{xy} < \Max(\alpha_{xy}),~~~
\Min(\alpha_{xz}) < \alpha_{xz} < \Max(\alpha_{xz}),
\eeq
where
\bey
\Min(\alpha_{xy})&=&
{1 \over 2} \arctan \left[{2c_{12} \over \Max(\gamma)+c_{11}-c_{22}} \right] 
\sim 25^o,\\\nonumber
~~~\Max(\alpha_{xy})&=&
{\pi \over 2} - {1 \over 2} \arctan \left[{2c_{12} \over c_{22}-c_{11}-\Min(\gamma)} \right] 
\sim 65^o,
\eey
\bey
\Min(\alpha_{xz})&=&
{1 \over 2} \arctan \left[{2c_{31}\over \Max(\gamma)+c_{11}-c_{33}} \right] 
\sim 7^o,\\\nonumber
~~~\Max(\alpha_{xz})&=&
{\pi \over 2} - {1 \over 2} \arctan \left[{2c_{31} \over c_{33}-c_{11}-\Min(\gamma)} \right]
\sim 65^o.
\eey
Necessary (but not sufficient) limits can also be derived analytically using, 
for example, conditions eq.~(\ref{alphaxylim}), and~(\ref{alphaxzlim}).
Upon substitution of eqs.~(\ref{j0}) and~(\ref{i0}) for our Gaussian triaxial 
models we find 
\bey
\alpha_{xy}&>&{1 \over 2} \arctan\left({2c_{12}J_0 \over I_0}\right)
={1 \over 2} \arctan \left({ 2c_{12} \over 2\sqrt{2}+c_{11} }\right)
\sim 10^o,~~~\\\nonumber
\alpha_{xz}&>&{1 \over 2} \arctan\left({2c_{31}J_0 \over I_0}\right)
={1 \over 2} \arctan \left({ 2c_{31} \over 2\sqrt{2}+c_{11} }\right)
\sim 3^o.
\eey
And likewise eq.~(\ref{xmass}) requires
\beq\label{fd}
\gamma \left(1-{3a^2 \over 2D_0^2}\right) + \left(2\sqrt{2}+{3 \over 2}c_{11}\right) >0.
\eeq
For an observer at $D_0 =4a$ this reduces to a lower limit on the amount
of phantom spheroid with $\gamma>-1.75$, or to an upper limit on the model
major axis angles with
\bey
\alpha_{xy} &=& {\pi \over 2} - {1 \over 2} \arctan \left({2c_{12} \over 
c_{22}-c_{11}-\gamma} \right) < 80^o,~~~\\\nonumber
\alpha_{xz} &=& {\pi \over 2} - {1 \over 2} \arctan \left({2c_{31} \over 
c_{33}-c_{11}-\gamma} \right) < 85^o.
\eey

\begin{figure}
\epsfysize=12cm
\centerline{\epsfbox{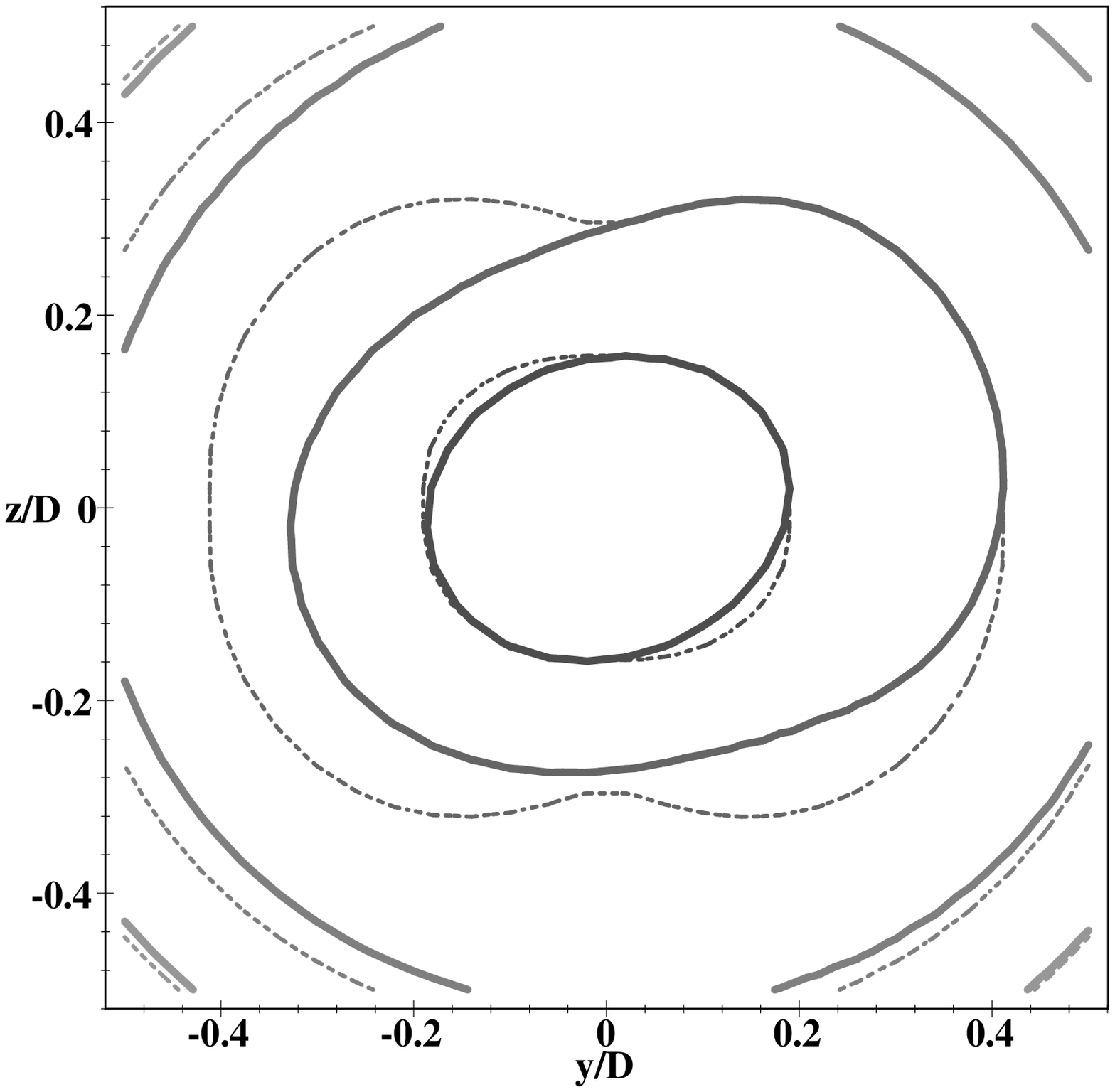}}
\caption{
The identical surface brightness map 
of the two triaxial models
from the observer's perspective, where the horizontal axis
$y/D$ and the vertical axis $z/D$ 
measure the angular distance of a line of sight
from the $y=0$ plane and the $z=0$ plane.  The dashed contours in the three
quadrants are the images of the contours in the first quadrant.
The offset with the solid contours shows the asymmetry or 
the perspective effect of a nearby object.  Contours
are in intervals of 1 magnitude.
}\label{barmap}
\end{figure}

\begin{figure}
\epsfysize=10cm
\centerline{\epsfbox{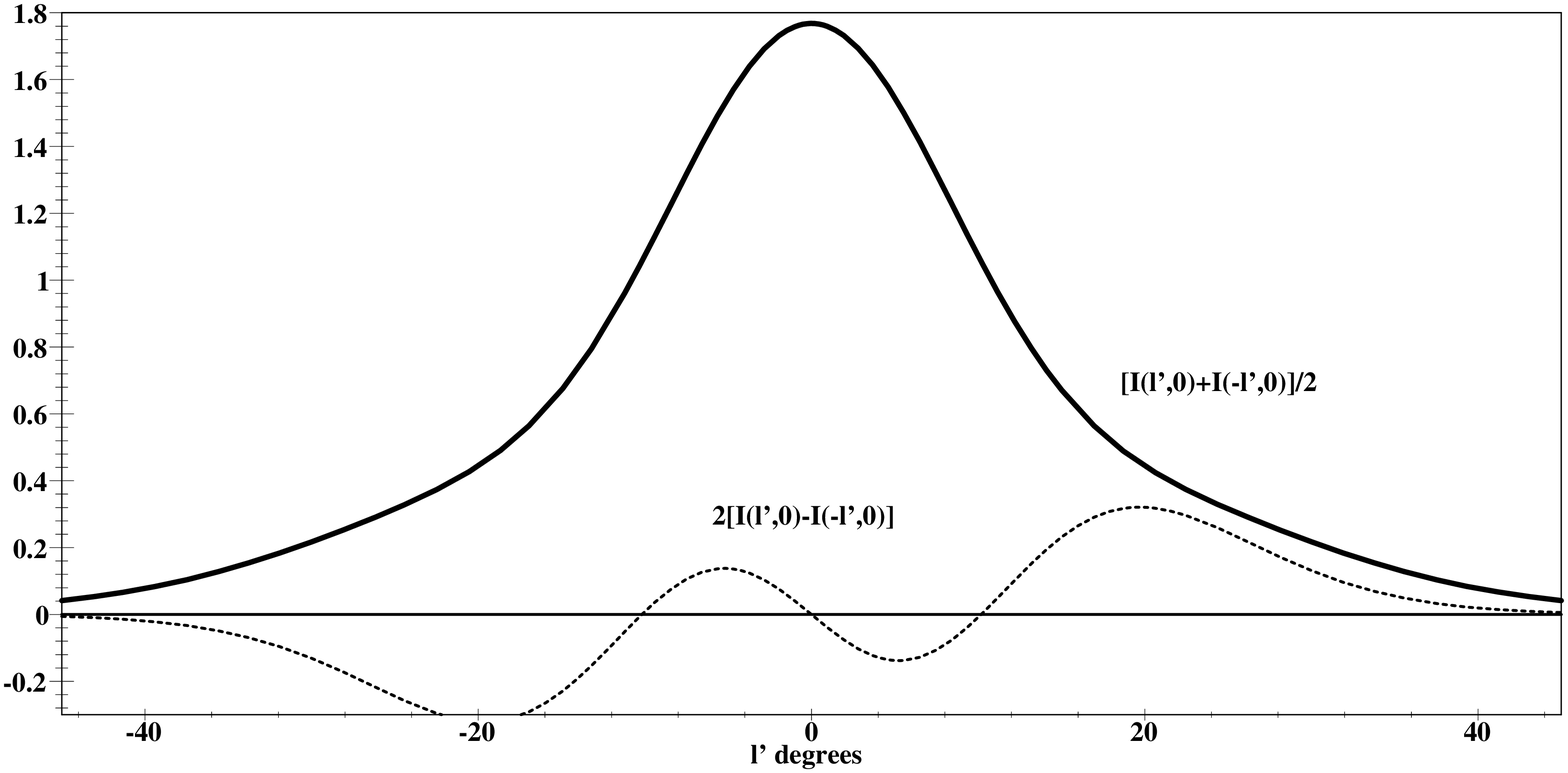}}
\vskip -2cm
\epsfysize=10cm
\centerline{\epsfbox{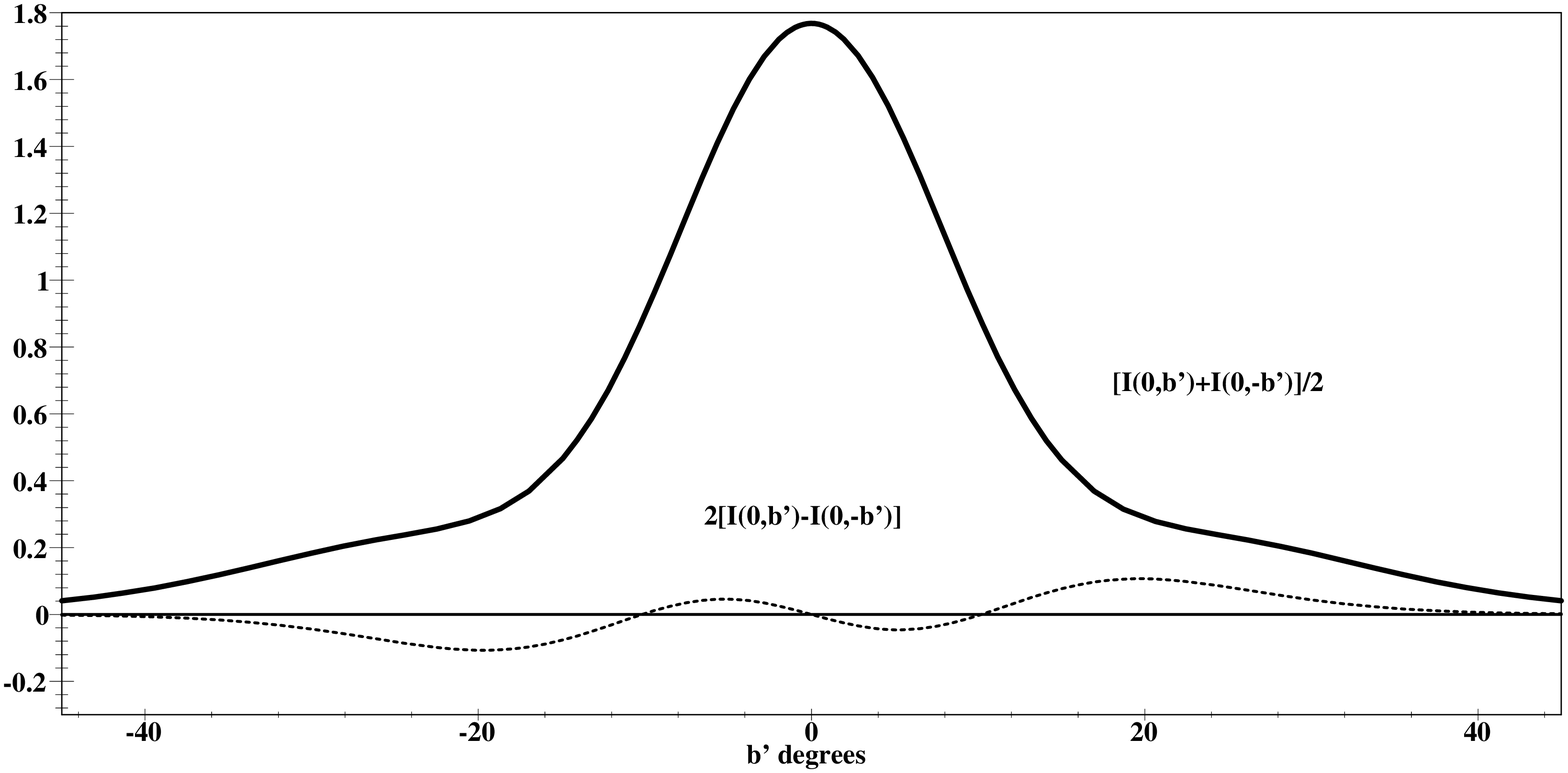}}
\caption{
the run of the surface brightness along the $l'$-axis (top)
and along the $b'$-axis (bottom) for our Gaussian triaxial models.
The solid curves show symmetric part and the dashed curves show
the perspective-induced asymmetry (after blown up by a factor of four).
The $l'$ and $b'$ are the generalized longitude and latitude
in units of degrees for the object.  
}\label{barlb}
\end{figure}

Fig.~\ref{bar1} and~\ref{bar2}
illustrate two Gaussian triaxial models, where
the parameter $\gamma$ is fixed by assigning the tilt 
angle $\alpha_{xy}$ (cf. eq.~\ref{alphaxyz}) 
in the $xy$ plane to $25^o$ or $50^o$.  
Figs.~(\ref{barmap}) and~(\ref{barlb}) show the identical 
surface brightness distribution of the two models, independent of $\gamma$.
Unlike the cigar-shaped bars in Fig.~\ref{cigar}, 
these two triaxial models are smooth with a positive density everywhere and 
an overall Gaussian radial profile except
for a small bump near ${D_0\over 2}$ (cf. Fig.~\ref{barxyz}).  

\begin{figure}
\vskip -1cm
\epsfysize=10cm
\centerline{\epsfbox{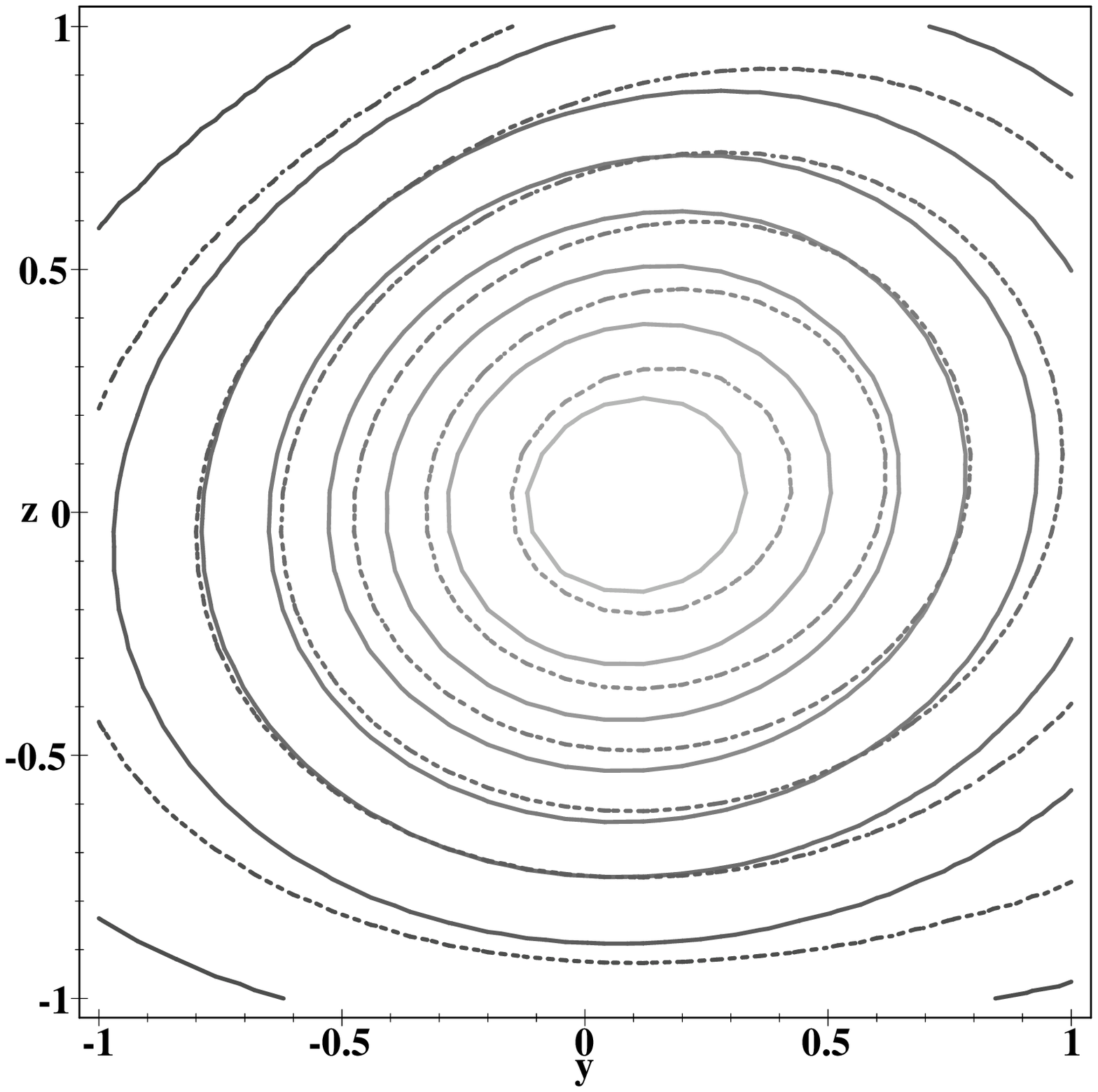}}
\vskip -2cm
\epsfysize=10cm
\centerline{\epsfbox{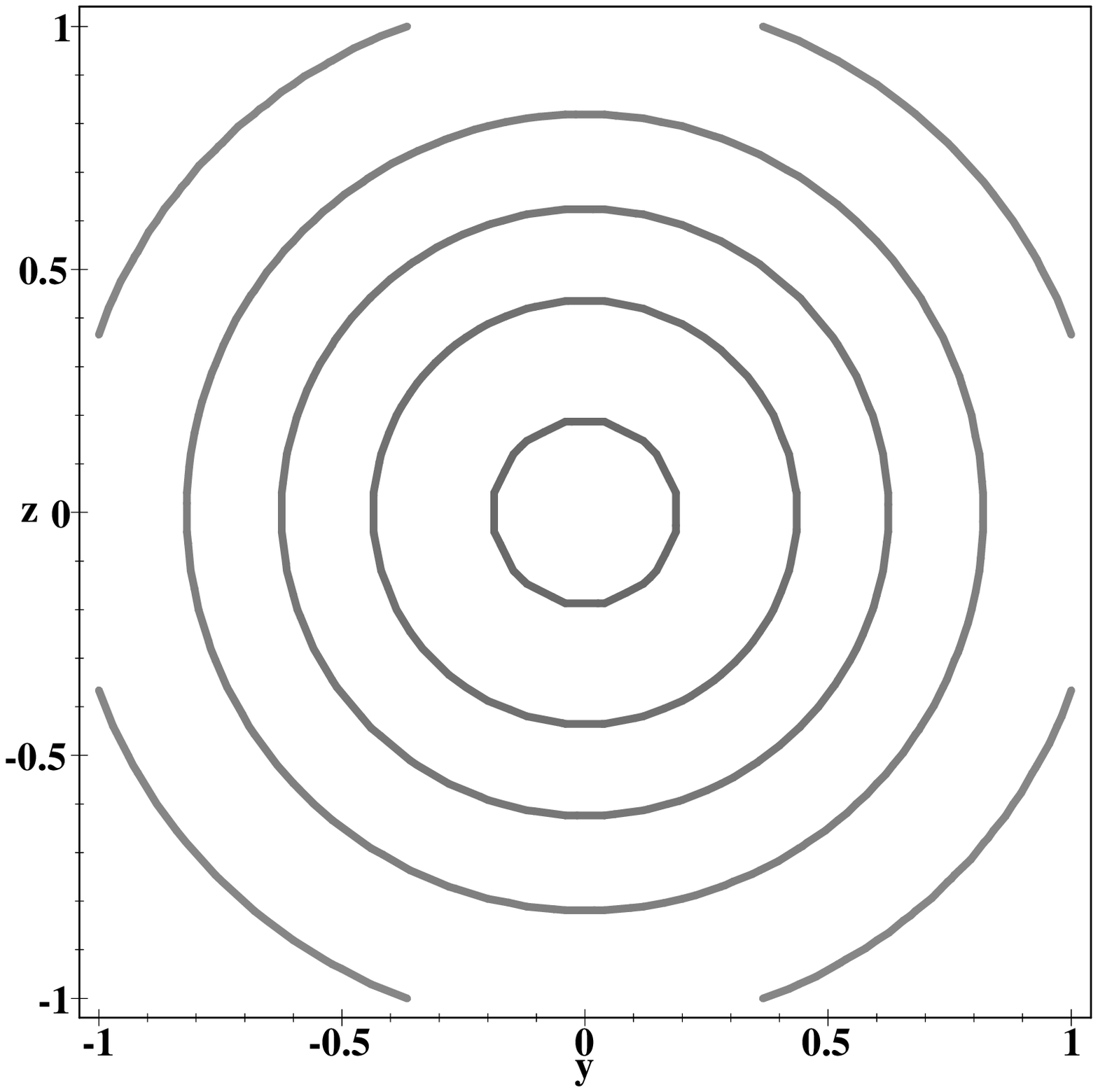}}
\caption{The cross-sections of the volume density at $x=2a/5$ for the 
$\alpha_{xy}=25^o$ and $\alpha_{xy}=50^o$ models (solid and dashed contours 
in the upper panel).  The maxima of the contours are where the
the long-axis of the bar intersects with the $x=2a/5$ plane.
They are offset from the $y=z=0$ point because the long-axis is
pointed away from the line of sight.
Subtracting one from another we get
the phantom spheroidal distribution (lower panel), which
has rotational symmetry around the $x$-axis.
}\label{slices}
\end{figure}
\begin{figure}
\vskip -1cm
\epsfysize=10cm
\centerline{\epsfbox{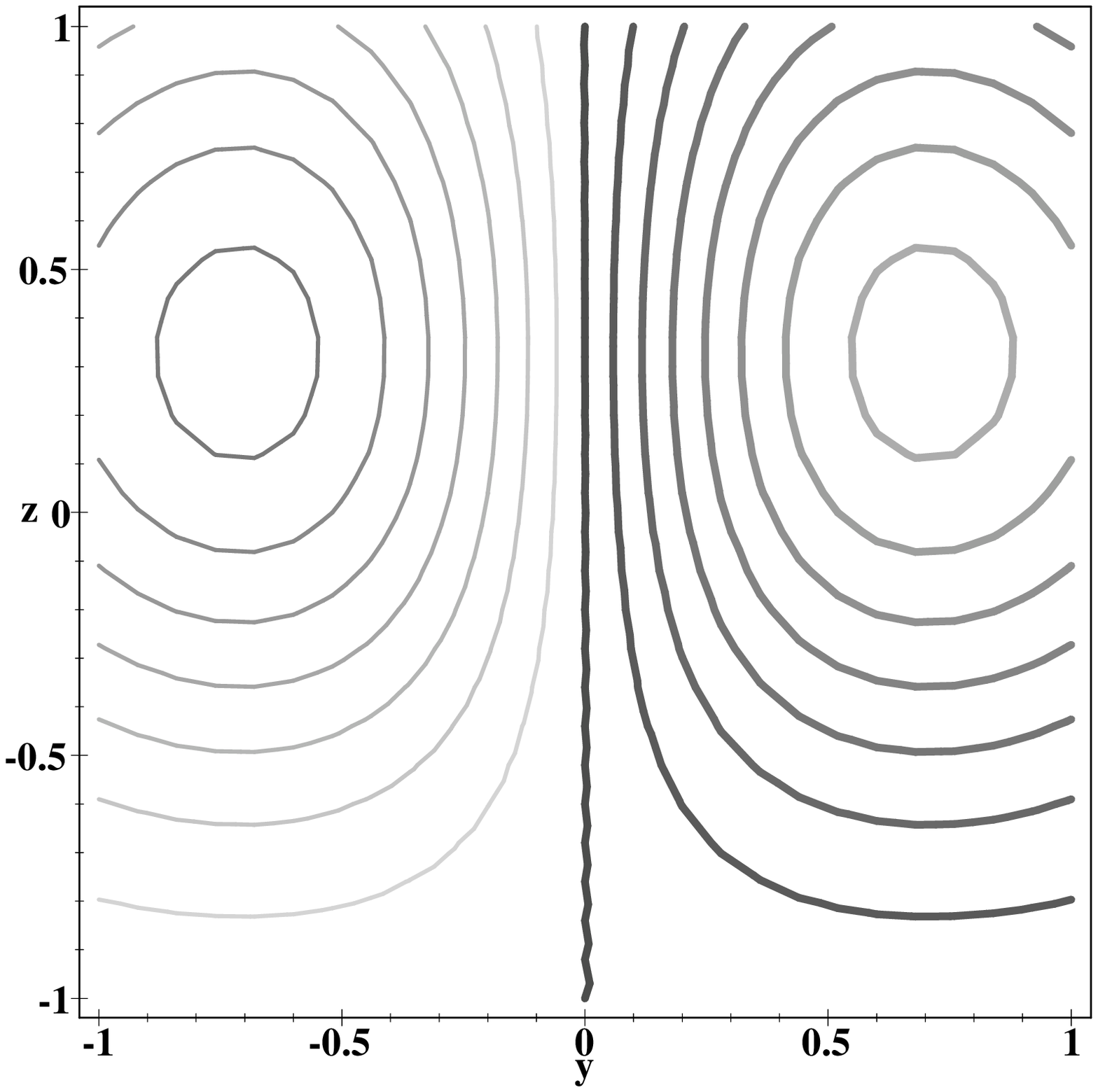}}
\vskip -2cm
\epsfysize=10cm
\centerline{\epsfbox{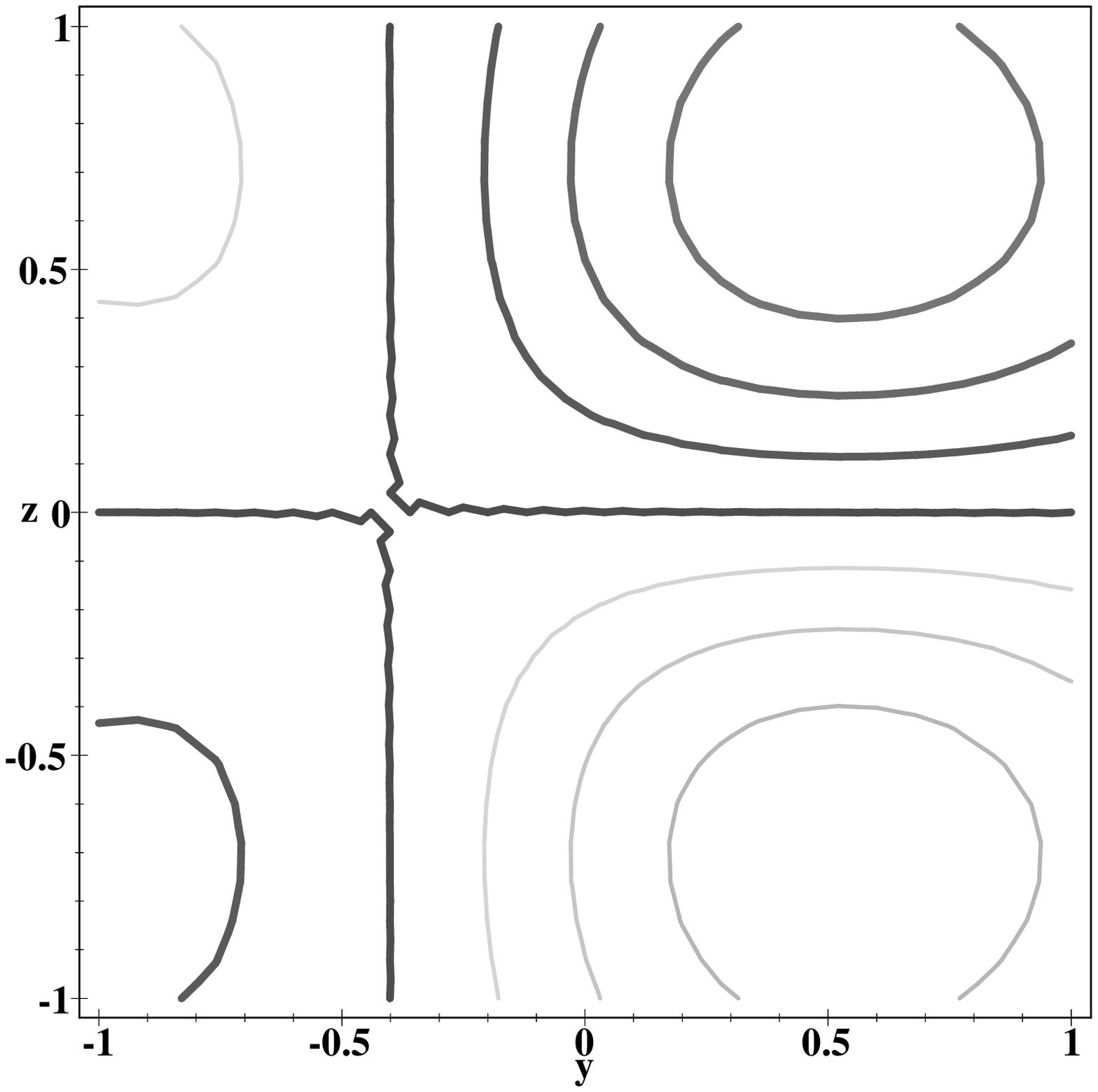}}
\caption{The cross-sections at $x=2a/5$ for the odd part of the model density,
which is independent of $\gamma$.  The upper panels shows 
$\nu_\gamma(x,y,z)-\nu_\gamma(x,-y,z)$, and lower panel shows
$\nu_\gamma(x,y,z)-\nu_\gamma(x,y,-z)$.  The distributions
are anti-symmetric with respect to the $y=l'=0$ plane
or the $z=b'=0$ plane with both positive (thick contours) 
and negative contours (thin contours).
}\label{sliceodd}
\end{figure}

The two models differ only in the even part of the volume density as
shown by the cross-sections at, say, $x=2a/5$ (cf. Fig.~\ref{slices}).
Subtracting one model from another we get back the phantom spheroid,
which is rotationally symmetric around the Sun-object center axis.

The odd part of the volume density is identical for both models,
and is shown in Fig.~(\ref{sliceodd}).
Mathematically, the odd part of the density is defined by
\beq\label{oddy}
\nu^{odd,y} \equiv {1 \over 2}\left[\nu_\gamma(x,y,z)-\nu_\gamma(x,-y,z)\right] = 
2y\left(c_{12}x+c_{23}z\right) p(r),
\eeq
and
\beq\label{oddz}
\nu^{odd,z} \equiv {1 \over 2}\left[\nu_\gamma(x,y,z)-\nu_\gamma(x,y,-z)\right] = 
2z\left(c_{13}x+c_{23}y\right) p(r).
\eeq
Clearly both $\nu^{odd,y}$ and $\nu^{odd,z}$ are independent of $\gamma$,
i.e., the amount of PS in the model.

The odd part of the density 
shows up as the asymmetry in the surface brightness.
The left-to-right asymmetry is a line-of-sight integration of $\nu^{odd,y}$ with
\beq\label{oddl}
{1 \over 2} \left[I(l',b')-I(-l',b')\right] 
= \int_{-\infty}^\infty  \! \! \nu^{odd,y} dD,
\eeq
and the up-to-down asymmetry is an integration of $\nu^{odd,z}$ along the
line of sight with
\beq\label{oddb}
{1 \over 2} \left[I(l',b')-I(l',-b')\right] 
= \int_{-\infty}^\infty  \! \! \nu^{odd,z} dD.
\eeq
More specific for the Gaussian triaxial models here, 
the asymmetry along the $b'=0$ cut is given by
\beq\label{etal}
{I(l',0)-I(-l',0) \over 2}
= I_0 \tilde{c}_{12} W(w) \sin 2l',
~~~\tilde{c}_{12} \equiv {c_{12} \over 2\sqrt{2}+c_{11} },
\eeq
and the asymmetry along the $l'=0$ cut is given by
\beq\label{etab}
{I(0,b')-I(0,-b') \over 2}
= I_0 \tilde{c}_{13} W(w) \sin 2b',
~~~\tilde{c}_{13} \equiv {c_{13} \over 2\sqrt{2}+c_{11} }
\eeq
where $\tilde{c}_{12}$ and $\tilde{c}_{13}$ are the combinations
of the intrinsic shape parameters of the model, and
\beq\label{eta}
W(w)=e^{-{w^2 \over a^2}} \left({2w^2 \over a^2}-1\right),
\eeq
is a function of the impact parameter $w$.  
Effectively $W(w)$ is a rescaled asymmetry distribution
after removing the periodical anti-symmetric term $\sin 2l'$ or $\sin 2b'$.
Note that the asymmetry distribution has a reversal of sign 
around the impact parameter 
\beq
w=a/\sqrt{2}.
\eeq
This means we can derive the scale length of the model $a$ 
directly from the observable asymmetry map.
For example, for the models in Fig.~(\ref{barlb}),
the object is at a distance $D_0=4a$,
so the reversal of the asymmetry, i.e., $W(w)=0$, happens at
\beq
|l'| = \arcsin{w \over D_0} \sim 10^o,
\eeq
in the longitudinal cut, or at $|b'| \sim 10^o$ in the latitudinal cut.
While fitting the asymmetry map will tell us unambiguously
the scale length $a$ and the shape parameter $\tilde{\epsilon}$
of the triaxial object, it does not directly constrain the major-axis 
angle $\alpha$,
nor does it constrain the amount of phantom spheroidal density.

\section{Phantom Spheroid vs. the Galactic bar}

This new form of non-uniqueness of nearby objects might apply to the
Galactic bar as well.  As mentioned in the Introduction, the
COBE/DIRBE surface brightness distribution is the main source of
information about the volume density of the Galactic bar.  Imaging the
effect of adding any amount of the phantom spheroids here $\gamma
F(x,y,z)$ to the COBE bar.  It does not matter whether we use the
cigar-shaped one in Fig.~\ref{cigar} or the smoother ones shown in
Fig.~\ref{inv}, as long as the PS is exactly centered at the Galactic
center, i.e., we set $D_0$ of these PS densities to the Galactocentric
distance (say 8 kpc).  The result is a new volume density of the
Galaxy with a generally twisted shape in the central part, but there
should be no trace of alteration in the the surface brightness maps
(the symmetric maps or the asymmetry maps).  This is purely because of
the construction of the PS (cf. eq.~\ref{FF}).  So the perspective
effect in the COBE/DIRBE maps is bypassed completely.  However, adding
any phantom spheroid here will modify only the even part of the volume
density and will have no influence on the anti-symmetric part since
the PS is always symmetric with respect to the $l=0$ plane and the
$b=0$ plane (cf. eq.~\ref{Ffold} and Fig.~\ref{slices}).  So as
illustrated by Fig.~(\ref{sliceodd}) and Fig.~(\ref{barmap}), the
prominent perspective effect in the left-to-right asymmetry map
$\left[I(l,b)-I(-l,b)\right]/2$ or the up-to-down asymmetry map
$\left[I(l,b)-I(l,-b)\right]/2$ should tightly constrain the part of
the volume density which is anti-symmetric with respect to the $l=0^o$
plane or the $b=0^o$ plane (cf. eqs.~\ref{oddl} and~\ref{oddb}).

It is tempting to suggest that phantom spheroids are responsible for
the uncertain major axis angle of the Galactic bar from the COBE/DIRBE map
(Binney \& Gerhard 1996, Binney, Gerhard \& Spergel 1997).  However,
this statement depends on whether there exists a phantom spheroidal
density which preserves the assumed mirror symmetry of the COBE bar
exactly.  Unless the added PS has a density profile matching that of
the COBE bar, it will create a $m=2$ twist, like a two-armed spiral
pattern or an S-shaped warp.  On the other hand, part of such
distortion may be hidden by the spiral arms in the Galactic disk.
After all the mirror symmetry is merely an assumption, which is
motivated by the fact that external face-on barred galaxies have a
largely bi-symmetric distribution of light.  Here we remark that the
sequence of Gaussian triaxial models here capture the main features of
the Galactic bar and the COBE/DIRBE maps: the models shown in
Fig.~\ref{bar1} and~\ref{bar2} with the scalelength $a=D_0/4=2$kpc
resemble the Gaussian Galactic bar models of Dwek et al. (1995) and
Freudenreich (1998) in terms of aspect ratios, radial profile and the
offset of the Sun from all three symmetry planes of the bar; the
asymmetric patterns in the surface brightness cuts in
Fig.~(\ref{barlb}) also mimics those seen in the COBE/DIRBE maps.
Hence it would not be surprising if there exists a phantom spheroid
which is qualitatively similar the one prescribed by eq.~(\ref{PS})
and preserves the mirror-symmetry of the Galactic bar.  Nevertheless a
quantitative determination of the parameters of the COBE bar is a
complex numerical problem involving to the least a reliable treatment
of dust extinction (Arendt et al. 1994, Spergel 1997).  Note that
patchy dust in the Galaxy makes a phantom spheroid here project to a
non-zero irregular surface brightness map.  The effect is strongest
when the dust is mixed with stars near the center, but even a dust
screen near the Sun can cast a faint shadow unless the phantom
spheroid is confined inside the solar circle (e.g. the cigar-shaped PS
in Fig.~\ref{cigar}).  The fact that dust extinction is a function of
wavelength could constrain the phantom spheroids.  The range of the
bar angle and axis ratio are also constrained by other observational
constraints of the bar such as microlensing optical depth (Zhao \& Mao
1996, Bissantz, Englmaier, Binney, \& Gerhard 1997) and star counts
(Stanek et al. 1994).  Star count data, for example, can constrain the
line-of-sight distribution of the bar, hence can in principle
distinguish between a prolate bar and a spherical bulge, and break the
degeneracy of the PS.  A detailed numerical treatment of the Galactic
bar is beyond the scope of this paper on the existence of
non-uniqueness in deprojecting a general nearby object.

\section{Phantom Spheroid vs. known non-uniqueness in external galaxies}

The phantom spheroidal density for nearby systems
is of a different nature from the kind of
non-uniqueness associated with a simple shear and/or stretch
transformation of an ellipsoid.  This applies even in the limit that
the object is at infinity because these transformations normally
change the odd part of the density, while the PS density here
is strictly an even function (cf. eq.~\ref{Ffold}).

Our phantom spheroidal density 
$F(x,y,z)$ is a function of the object distance $D_0$.  
In general
\beq\label{PSD0}
F(x,y,z)=F_{D_0\rightarrow\infty} (x,y,z) + D_0^{-2} F_0(r),
\eeq
where 
$F_{D_0\rightarrow\infty}$ is the PS in the limit of 
$D_0 \rightarrow \infty$, and $F_0(r)$ is (up to a constant) the PS 
in the limit of $D_0 \rightarrow 0$.   For
a hypothetical observer receding from the object, the distance $D_0$
increases, the PS density is adjusted slightly to accommodate
the changing perspective so to preserve this kind of non-uniqueness
all the way to extragalactic distance.  The common ambiguity of an
extragalactic axisymmetric bulge with an end-on bar is a very special
case of the kind of non-uniqueness for nearby objects.

The divergence of $F(x,y,z)$ at small $D_0$
(cf. eq.~\ref{PSD0}) has an interesting implication.
In the limit that $D_0\rightarrow 0$
some regions of the phantom spheroid will have infinitely positive density,
and some regions infinitely negative density because
$\gamma F(x,y,z) \rightarrow  \gamma D_0^{-2} F_0(r)$, and $F_0(r)$ changes sign,
for example, at $r=a\sqrt{3/2}$ (cf. eq.~\ref{PS} or Fig.~\ref{invplts}).
So to add any finite amount of PS would violate the positivity requirement.  
Hence only the model with $\gamma=0$ is allowed, 
and the major axis direction is fixed to the direction where the
object appears the brightest in the projected map.
In other words the kind of non-uniqueness
discussed here disappears if the observer is very close to the object center.
In general the range for $\gamma$ becomes very narrow if $D_0 \le a$, and 
is insensitive to the distance $D_0$ to the object if $D_0>2a$.  

Unlike its external counterpart with $D_0 \rightarrow \infty$,
the PS density for a nearby triaxial body is not massless.
The total luminosity $L_\gamma$ of our model 
changes with $\gamma$.  For the Gaussian model we have 
(cf. eqs.~\ref{Lu} and~\ref{nu2})
\bey\label{lum}
L_\gamma&=& \int\!\! d^3{\bf r} \nu_\gamma(x,y,z) = 
L_0 + \gamma L_{PS},\\\nonumber
L_0 &=& \pi^{3 \over 2} \nu_c a^3 \left[2^{3 \over 2} + {c_{11}+c_{22}+c_{33} \over 2}\right],\\\nonumber
L_{PS}&=&-\left({3 \pi^{3 \over 2} \nu_c a^3 \over 2}\right) 
\left({\gamma a^2 \over D_0^2}\right),
\eey
where $L_0$ is the luminosity in the absence of 
the PS.  The PS has a luminosity $L_{PS}$ proportional to $D_0^{-2}$.

The normalization $\nu_c$ and the scale $a$ are fixed by the projected
light intensity $I_0$ and the asymmetry map $W(w)$
(cf. eqs.~\ref{eta} and~\ref{i0}), but the total luminosity changes by
a fraction of the order of ${a^2 \over D_0^2}$.  This is because
unlike external systems, the observed angular distribution of light in
a nearby object does not simply sum up to a unique measurement of its
total luminosity.  We may slightly underestimate/overestimate the
intrinsic luminosity by about a few percent, which may not be easy to
detect in reality.  Moving away from the object the observer has a
full outside view and hence a better determination of its luminosity
but at the expense of relaxing the constraint on the orientation of
the object from the perspective effect and positivity.

The phantom spheroid affects also the dynamics of the model,
however, the effect can be fairly difficult to detect.
For the models in eq.~(\ref{nu}), the
depth of the potential well at the center expressed in terms of
the escape velocity $V_{esc}$ is given by
\beq
{1 \over 2} V_{esc}^2 = \pi \nu_c a^2 G(M/L) 
\left\{\left[ 4 + {2  \over 3} \left(c_{11}+c_{22}+c_{33}\right) \right]
- \left( {1 \over 3} + {a^2 \over D_0^2} \right) \gamma \right\},
\eeq
which decreases linearly with increasing fraction ($\gamma$) of the superimposed 
PS, where $M/L$ is the mass-to-light ratio.
As a result, the potential well of the $\alpha=25^o$ model 
is slightly shallower 
than that of the more centrally concentrated $\alpha=50^o$ model 
(cf. Fig.~\ref{bar1} and~\ref{bar2}).
However, the difference is only 7\% in terms of the maximum escape velocity
$V_{esc}$ of the models.
The differences in terms of
the mass weighted average velocity dispersion (estimated from the virial
theorem) and the circular velocity (estimated directly from the potential) 
are also at only a few percent level, too small to be measured with certainty.
However it might still be feasible to distinguish the models in terms of 
the stellar orbits in them.  
The projected velocity distribution should change with the major axis angle.

\begin{figure}
\epsfysize=9cm
\vskip -1cm
\centerline{\epsfbox{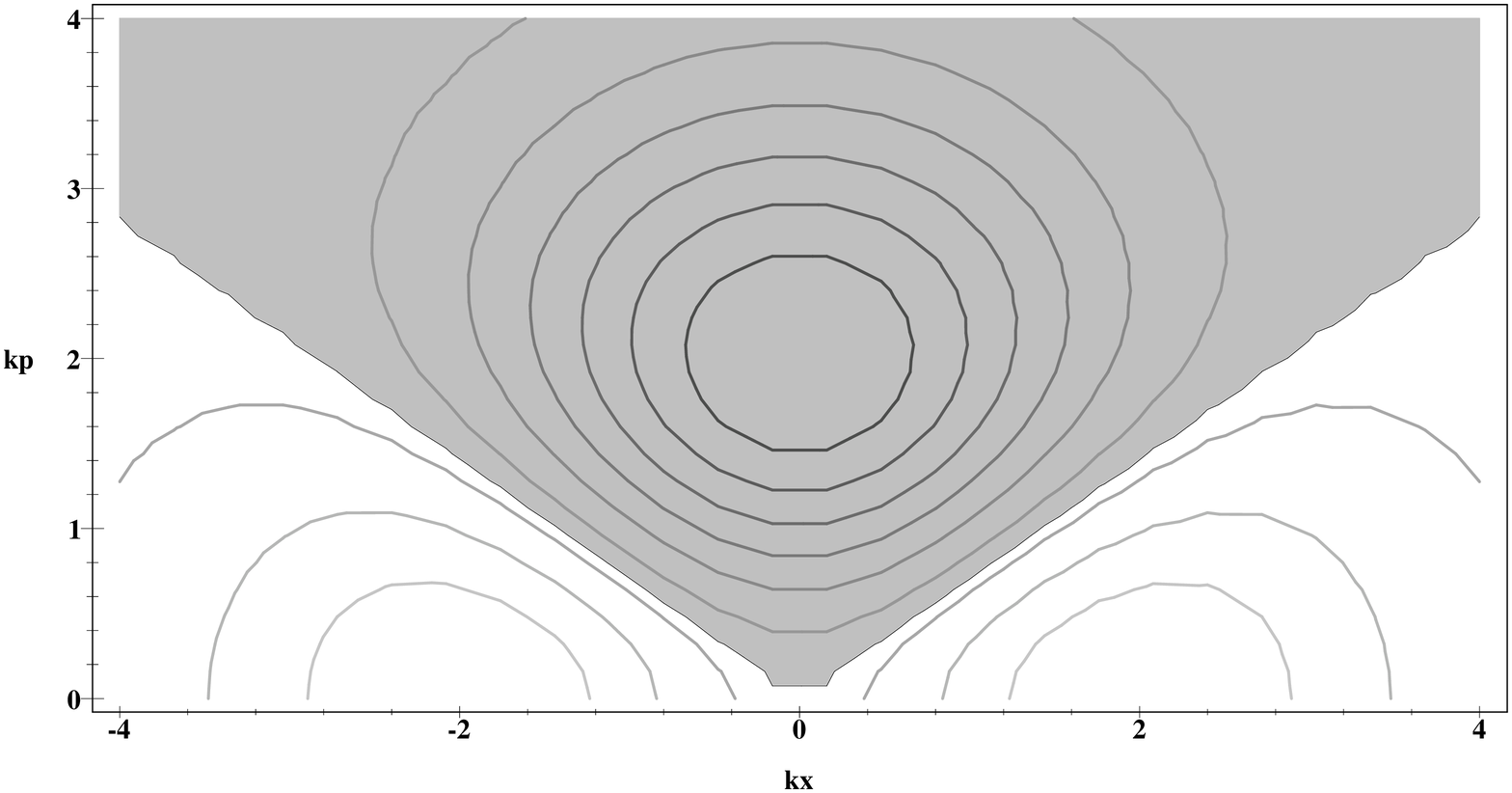}}
\vskip -1.5cm
\epsfysize=9cm
\centerline{\epsfbox{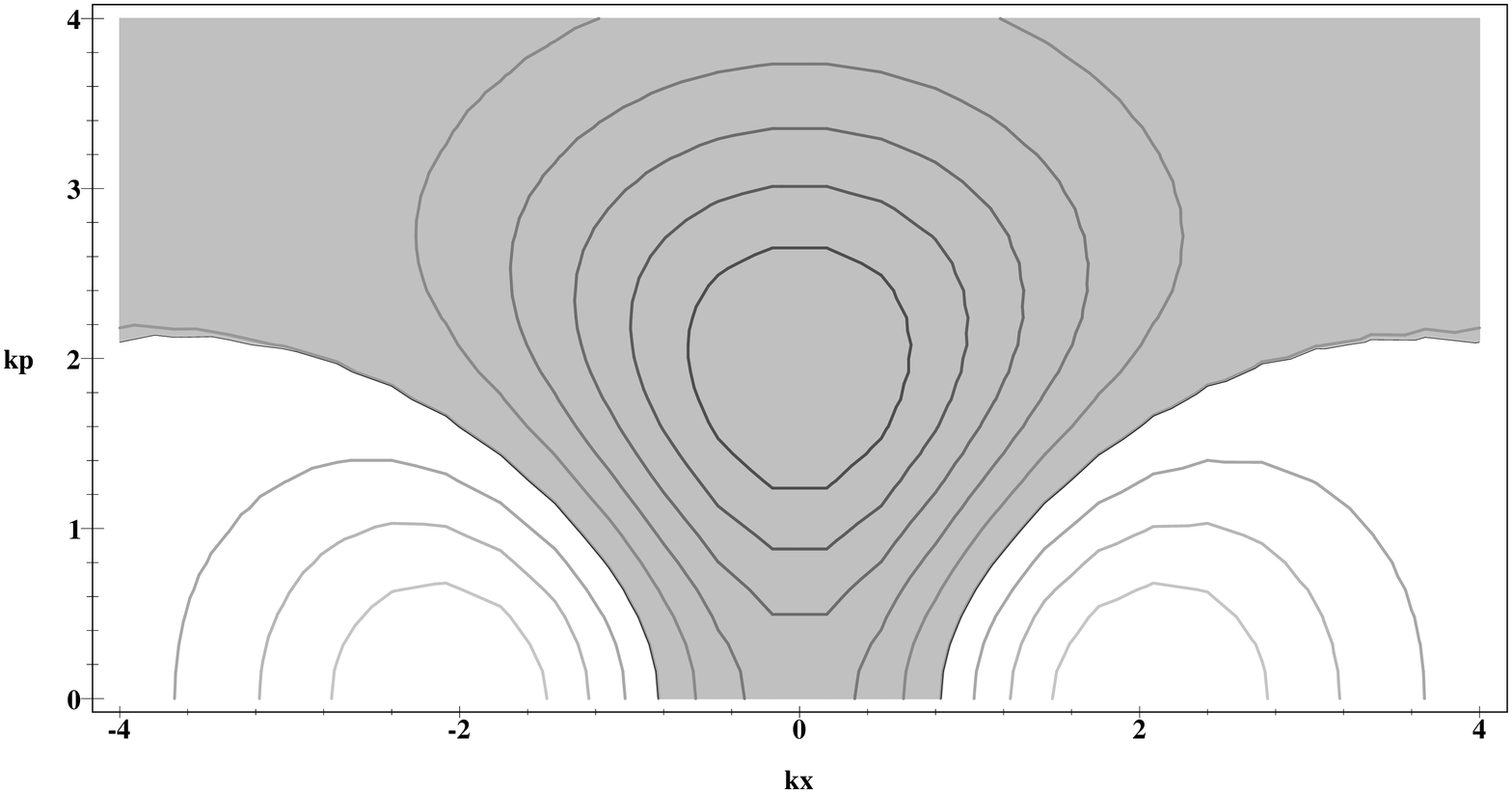}}
\vskip -1.5cm
\epsfysize=9cm
\centerline{\epsfbox{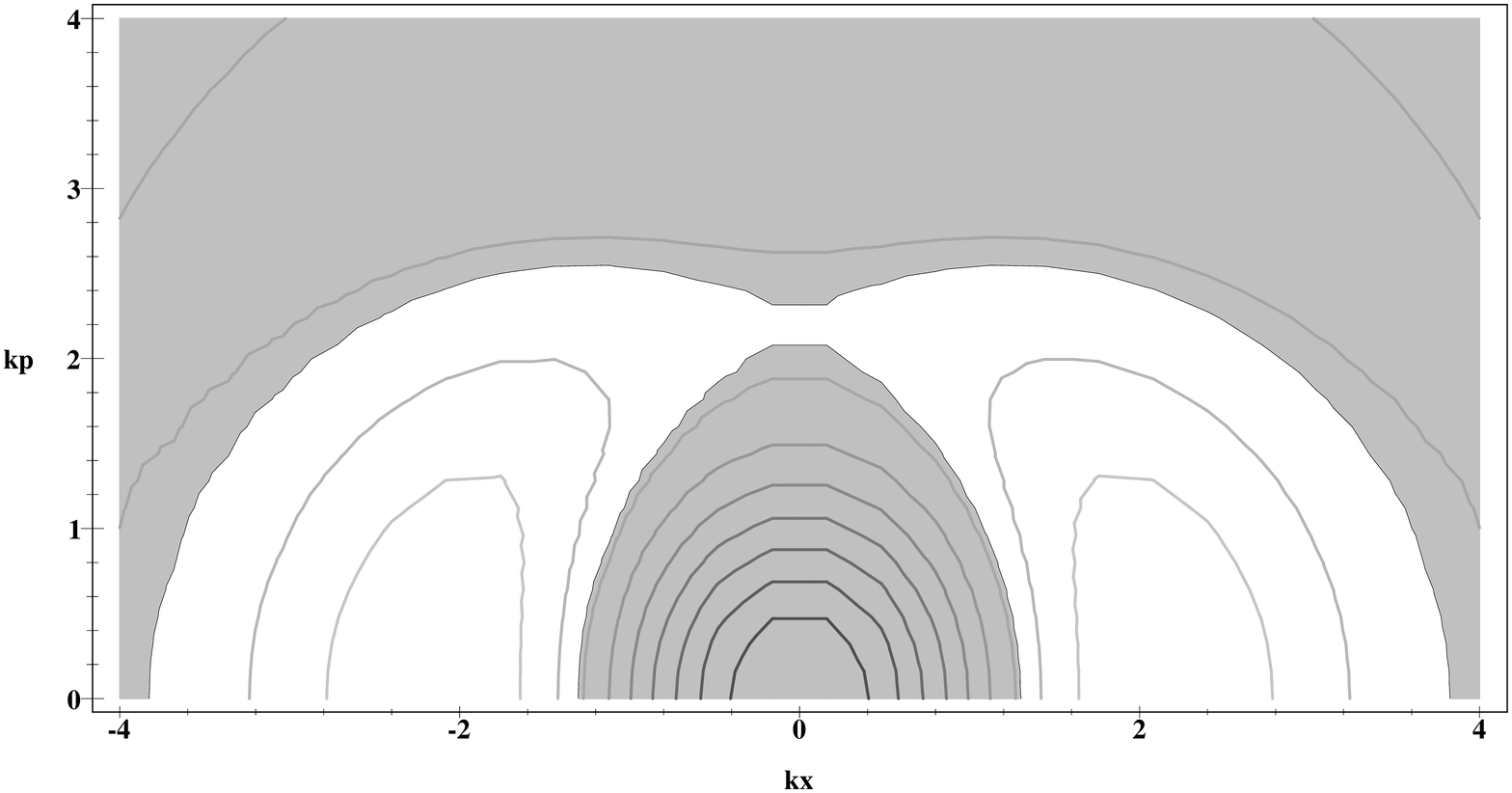}}
\vskip -1cm
\caption{
The phantom density in the Fourier ${\bf k}$-space, $k_x$ vs. 
$k_p=\sqrt{k_y^2+k_z^2}$ in units of $a^{-1}$.
The distributions are rotationally symmetric around the $k_x$-axis.
The shaded and unshaded zones correspond to negative and positive 
$\tilde{F}({\bf k})$ respectively.
The observer is at $D_0=\infty, 4a, 1.25a$ from top to bottom.  
}\label{fk}
\end{figure}
Our phantom spheroidal density is also different from konuses since 
it is not confined to any cone in the Fourier
{\bf k}-space even in the limit $D_0 \rightarrow \infty$.
Using eqs. (4), (24) and (25) of Palmer (1994) we can 
compute the Fourier transform of our PS (cf. eq.~\ref{PS}).  We find
\beq\label{kspace}
\tilde{F}({\bf k}) \equiv 
\int \! \! d^3{\bf r} \exp(-i {\bf k} \cdot {\bf r} )\left[p(r) x^2 -S(r)\right]
=\tilde{F}_{D_0 \rightarrow \infty} ({\bf k}) + D_0^{-2} \tilde{F}_0({\bf k}),
\eeq
which is a {\bf k}-space spheroidal distribution around the Sun-center line,
where
\beq\label{kinfty}
\tilde{F}_{D_0 \rightarrow \infty} ({\bf k})
\equiv \pi^{3 \over 2}  \nu_c a^3 
\left(\cos^2 \theta_{{\bf k}} -{2 \over 3}\right)
{k^2a^2 \over 4} e^{-{k^2a^2 \over 4}},
\eeq
$\theta_{{\bf k}}$ 
is the angle of the {\bf k}-vector with the Sun-center line,
and
\beq
D_0^{-2} \tilde{F}_0({\bf k})  \equiv  
\left(1 - {7 k^2a^2 \over 12} + {k^4a^4 \over 24} \right) 
e^{-{k^2a^2 \over 4}} L_{PS}.
\eeq
We recover the luminosity of the PS in the limit ${\bf k} \rightarrow 0$,
$\tilde{F}(0)=D_0^{-2}\tilde{F}_0(0)=L_{PS}$
(cf. eq.~\ref{lum}).  Fig.~\ref{fk} 
shows that even when the object is placed at infinity,
$\tilde{F}({\bf k})$ is non-zero everywhere except 
at {\bf k}$=0$ or $\theta_{{\bf k}}=\cos^{-1} \sqrt{2 \over 3}$
(cf. eq.~\ref{kinfty}).  
Since any konus-like structure (or its triaxial version) must have 
a certain ``cone of ignorance'' around a principal axis
of the model outside which $\tilde{F}({\bf k})=0$
(Rybicki 1986, Gerhard \& Binney 1996,
Kochanek \& Rybicki 1996), the kind of non-uniqueness shown in this paper
has little to do with konuses.  The two sequences of models meet
only when the object is a face-on or end-on spheroidal system, in which case, 
the PS here is equivalent to a trivial konus with
the entire ${\bf k}$-space in the ``cone of ignorance''.

\section{Conclusion}

The non-uniqueness in deprojecting the surface brightness maps of a
nearby system can originate from the simple fact that a spherical
bulge can match an end-on prolate bar or a face-on disk in the
line-of-sight integrated light distribution.  Subtracting the
spherical bulge from the prolate end-on bar or the face-on disk forms
a phantom spheroid (PS), which has both positive and negative density
regions.  The non-uniqueness due to such PS can be characterized by
two numbers: the distance ($D_0$) of the object and the amount
($\gamma$) of the PS density that we can superimpose.  By tailoring
the PS, it is possible to preserve the triaxial reflection symmetries
of the model.  The orientation of these symmetry planes with respect
to our line of sight changes as increasing amount of PS is added up to
the point when negative density regions appear in the final model.
The limits on the major-axis angles are given analytically.
The phantom spheroidal density here forms a new class of
non-uniqueness of entirely different nature from known degeneracy in
deprojecting extragalactic objects.  It does not preserve the total
luminosity of the system.

The author thanks Frank van den Bosch and Tim de Zeeuw for 
many helpful comments on the presentation, and Ortwin Gerhard
for careful reading of an early draft.

\appendix 

\section{Positivity as a constraint to the major axis angle of edge-on models}

Edge-on models are specified by
\beq
c_{23}=c_{31}=\alpha_{yz}=\alpha_{xz}=0,
~~~c_{12} \equiv \epsilon,~~~\alpha \equiv \alpha_{xy},
\eeq
so that the $z=0$ plane, 
which passes through the observer at $(x,y,z)=(D_0,0,0)$,
is a symmetry plane of the model.
The volume density model of the system is
best described in a cylindrical coordinate system $(R,Z,\psi)$
centered on the object.  This coordinate system is related to 
the rectangular coordinate system by
\beq
x=R \cos\psi, ~~~y=R \sin\psi,~~~z=Z,
\eeq
where the $Z$-axis is the symmetry axis of the edge-on system,
and $\psi=0$ is a plane, passing through the Sun-object line and the $Z$-axis.
In this coordinate system we can rewrite the ellipsoidal term $Q_\gamma$ 
(cf. eq.~\ref{Qu}) as
\beq
Q_\gamma = \left[ c_{33}Z^2 + \left(c_{22}+\epsilon\cot 2\alpha\right)R^2 \right] + 
{\epsilon R^2 \over \sin 2\alpha} \cos(2\psi-2\alpha).
\eeq
Clearly surfaces of constant $Q_\gamma$
are ellipsoidal surfaces with mirror symmetry
with respect to the $\psi=\alpha$ plane, the $\psi=\alpha+{\pi \over 2}$ plane
and the $Z=0$ plane.  

The triaxial model density (cf. eq.~\ref{nu1}) 
can then be written in above notations as
\beq\label{nu3}
\nu_\gamma(R,Z,\psi) = n_0(R,Z) + n_2(R,Z) \cos(2\psi-2\alpha),
\eeq
where the second term is a triaxial perturbation
with the major axis along $\psi=\alpha$ and amplitude
\beq
n_2(R,Z) ={\epsilon R^2p \over \sin 2\alpha},
\eeq
and the first term, the axisymmetric part, is given by
\beq
n_0(R,Z) \equiv E(R,Z) + \epsilon \cot 2\alpha \left[R^2p-2S\right],
\eeq
where
\beq
E(R,Z) \equiv \left[G + (c_{11}-c_{22}) S\right] + (c_{22}R^2+c_{33}Z^2)p.
\eeq

Imposing positivity requirements to $\nu_\gamma(R,Z,\psi)$ for all values
of the azimuthal angle $\psi$, we have 
\beq
n_0(R,Z) \ge n_2(R,Z),
\eeq
i.e.,
\beq
E(R,Z) + \epsilon \cot 2\alpha \left[R^2p-2S\right] 
\ge {\epsilon R^2p \over \sin 2\alpha}.
\eeq
Using the following relations for sinusoidal functions
\beq
{1 \over \sin 2\alpha} = {1+t^2 \over 2t},~~\cot2\alpha = {1-t^2 \over 2t},
~~t \equiv \tan \alpha, 
\eeq
the inequality reduces to an upper and a lower bound for 
the angle $\alpha$ with, 
\beq\label{arange}
\Max\left[t_{-}\right] \le \tan \alpha \le \Min\left[t_{+}\right],
\eeq
where $t_{-} \le t \le t_{+}$ is the range
bounded by the effectively quadratic inequality for $t$
\beq\label{tpm}
\left[\epsilon R^2p - \epsilon S\right] t + \epsilon S t^{-1} \le E(R,Z)
\eeq
at a given position $(R,Z)$ on the meridional plane,
and the overlapped interval of these ranges is used in eq.~(\ref{arange}).  
For example, for models with
parameters in eq.~(\ref{examp}) and $D_0=4a$, we find $77^o > \alpha > 15^o$
to ensure a positive density at $(R,Z)=(a,a)$, but a narrower range
$63.5^o \ge \alpha \ge 22.5^o$ is necessary to ensure positivity
everywhere in the $(R,Z)$ plane. 

Interestingly we recover the results of eq.~(\ref{alphaxylim})
if we integrate both sides of eq.~(\ref{tpm}) along the $y$-axis.  
And likewise if we multiply 
both sides of eq.~(\ref{tpm}) by a factor $x^2$ and then 
integrate along the $x$-axis, we recover the results of eq.~(\ref{xmass}).

\section{Analytical expressions for a series of phantom spheroids}

New phantom spheroids can be generated from the original PS $F(x,y,z)=x^2p(r)-S(r)$
(cf. eq.~\ref{PS}) by repeatedly applying the derivative operator 
${1 \over 2}{\partial \over \partial \ln a}$, i.e,
\beq\label{PSNEW1F}
F_1(x,y,z) ={\partial F(x,y,z) \over 2 \partial(\ln a)}
=x^2p_1(r)-S_1(r),
\eeq
where
\beq\label{PSNEW1p}
p_1(r) \equiv {\partial p(r) \over 2 \partial(\ln a)}
=-p(r)\left(1-{r^2 \over a^2}\right),
\eeq
\beq\label{PSNEW1S}
S_1(r) \equiv {\partial S(r) \over 2 \partial(\ln a)}
=a^2p(r)\left[{r^2 \over 2 a^2} + {a^2 \over D_0^2} 
\left(-{5r^4 \over 2a^4}+{r^6 \over a^6} \right) \right],
\eeq
and
\beq\label{PSNEW2F}
F_2(x,y,z) ={\partial F_1(x,y,z) \over 2 \partial(\ln a)}
=x^2p_2(r)-S_2(r),
\eeq
where
\beq\label{PSNEW2p}
p_2(r) \equiv {\partial p_1(r) \over 2 \partial(\ln a)}
=p(r)\left(1-{3r^2 \over a^2}+{r^4 \over a^4}\right),
\eeq
\beq\label{PSNEW2S}
S_2(r) \equiv {\partial S_1(r) \over 2 \partial(\ln a)}
=-a^2p(r)\left[\left({r^2 \over 2 a^2}-{r^4 \over 2 a^4}\right)
+ {a^2 \over D_0^2} 
\left(-{5r^4 \over 2a^4}+{9r^6 \over a^6}-{r^8 \over a^8} \right) \right].
\eeq


{}

\bsp
\label{lastpage}
\end{document}